\documentclass[prl,twocolumn]{revtex4-1}

\usepackage{natbib}
\usepackage{graphicx}
\usepackage{amssymb}
\usepackage{amsmath}
\usepackage{bm}

\newcommand{\ave}[1]{\ensuremath{\left\langle #1\right\rangle}}
\newcommand{\Ham}{\mathcal{H}}
\newcommand{\J}{\mathcal{J}}
\newcommand{\bra}[1]{\ensuremath{\left\langle #1\right|}}
\newcommand{\ket}[1]{\ensuremath{\left| #1\right\rangle}}
\renewcommand{\vec}[1]{\mathbf{#1}}

\begin{document}

\title{On the ferromagnetic ground state of SmN}

\author{J.~F.~McNulty}
\author{B.~J.~Ruck}
\author{H.~J.~Trodahl}

\affiliation{The MacDiarmid Institute for Advanced Materials and
  Nanotechnology, School of Chemical and Physical Sciences, Victoria
  University of Wellington, P.O. Box 600, Wellington, New Zealand}

\begin{abstract}
  SmN is a ferromagnetic semiconductor with the unusual
  property of an orbital-dominant magnetic moment that is
  largely cancelled by an antiparallel spin contribution,
  resulting in a near-zero net moment.  However, there is a
  basic gap in the understanding of the ferromagnetic ground
  state, with existing density functional theory calculations
  providing values of the $4f$ magnetic moments at odds with
  experimental data. To clarify the situation, we employ an
  effective $4f$ Hamiltonian incorporating spin-orbit
  coupling, exchange, the crystal field, and $J$-mixing to
  calculate the ground state $4f$ moments. Our results are
  in excellent agreement with experimental data, revealing
  moderate quenching of both spin and orbital moments to
  magnitudes of $\sim 2~\mu_B$ in bulk SmN, enhanced to an
  average of $\sim 3~\mu_B$ in SmN layers within a SmN/GdN
  superlattice. These calculations provide insight into
  recent studies of SmN showing that it is an unconventional
  superconductor at low temperatures and displays twisted
  magnetization phases in magnetic heterostructures.
\end{abstract}

\date{\today}
\maketitle
\section{Introduction}
In recent years, research into SmN has been driven by two
main considerations; its potential for spintronic
applications and its vanishingly small ferromagnetic moment.
It is a member of the rare-earth nitride series, a group of
materials owing their promise for spintronic applications to
their status as intrinsic ferromagnetic semiconductors
\cite{Natali_PMS_2013,Leuenberger_PRB_2005,Granville_PRB_2006,
  Preston_PRB_2007,Meyer_PRB_2008,Meyer_JMMM_2010,Richter_PRB_2011,
  Binh_PRL_2013, Azeem_JAP_2013,Warring_PRB_2014}. SmN is
especially unique among the series, with a vanishingly small
ordered moment arising from the near-cancellation of spin
and orbital moments \cite{Preston_PRB_2007,Meyer_PRB_2008}
fixed antiparallel by strong spin-orbit coupling. So far the
spintronic potential of the RENs remains largely untapped,
though GdN has been integrated within spin-filters
\cite{Senapati_Nature_2011,Massarotti_NC_2015} and field
effect transistors \cite{Warring_APL_2013} while DyN has
been incorporated into a magnetic tunnel junction
\cite{Muduli_PRB_2014}. However, SmN has quite recently
displayed remarkable phenomena in magnetization and
transport studies. There is a report of an unconventional
twisted magnetization phase occurring in SmN films exchange
coupled to GdN, observed via x-ray magnetic circular
dichroism (XMCD)\cite{McNulty_PRB_2015}, and the discovery of
superconductivity coexisting with ferromagnetic order at
temperatures of $3$-$5$~K in SmN films and SmN/GdN
superlattices, thought to be due to a spin-triplet pairing
mechanism \cite{Anton_Arxiv_2015}. These phenomena are
intimately related to the ferromagnetic state of SmN, which
displays ferromagnetic order below $T_C \approx 27$~K with a
near-zero moment of 0.035$~\mu_B$ per Sm$^{3+}$, with the
unusual situation that the orbital moment is marginally
dominant, and defines the net magnetization direction
\cite{Preston_PRB_2007,Meyer_PRB_2008,Anton_PRB_2013}.

The magnetic state of SmN originates from the partially
filled $4f$ shell of the trivalent Sm ion, the usual valence
among the RENs except CeN and EuN
\cite{Natali_PMS_2013,Richter_PRB_2011,Ruck_PRB_2011,Binh_PRL_2013},
with Hund's rules providing the ground state
configuration. In Sm$^{3+}$ ($4f^5$), Hund's rules yield a
$4f$-shell spin alignment of $S=5/2$, with the spin-orbit
coupling enforcing an opposed orbital alignment of $L=5$,
resulting in a total angular momentum quantum number
$J=|L-S|=5/2$ and Land\'{e} factor of $g_J=2/7$.  The
$J=5/2$ ground state-multiplet of Sm$^{3+}$ is well-known to
be influenced by the first excited $J=7/2$ multiplet which
is separated by $\Delta E \approx 1500~$K from the ground
state
\cite{DeWijn_PSS_1976,Adachi_PRB_1997,Meyer_PRB_2008}. However,
the effect of this excited state is not large enough to
explain the discrepancy between the experimentally measured
SmN moment of $0.035~\mu_B$ and the theoretical free-ion
saturation moment of the ground-state multiplet
$\mu_B\ave{L_z+2S_z}_{GS}=\mu_Bg_JJ = 0.74~\mu_B$
\cite{Meyer_PRB_2008}. The experimental 0.45~$\mu_B$
paramagnetic moment is almost half the free-ion value of
$\mu_Bg_J\sqrt{J(J+1)}= 0.85~\mu_B$, but is close to the
0.41~$\mu_B$ moment expected for a $\Gamma_7$ crystal-field
ground state doublet \cite{Meyer_PRB_2008}, signalling that
the crystal field may also play an important role in the
small ferromagnetic moment. The low Curie temperature of
$\approx 27~$K in SmN fits in with the trend of low ordering
temperatures in the REN series \cite{Natali_PMS_2013} which
is due to weak indirect exchange. The highly localized $4f$
states ensure that the inter-ionic exchange mechanism
precipitating ferromagnetism proceeds via Sm $5d$ and N $2p$
states in indirect exchange processes
\cite{Mitra_PRB_2008,Sharma_PRB_2010,
  Natali_PMS_2013}. Because of the semiconducting nature of
the RENs, the carrier-mediated RKKY mechanism does not
appear to be the dominant exchange channel as it is in
metallic rare-earth systems
\cite{Natali_PRB_2013,Lee_APL_2015}.

To date there has been no convincing description of SmN
which accounts for both the magnitude and orbital-dominant
sign of the ordered moment. Furthermore, XMCD experiments on
bulk SmN and a SmN/GdN superlattice show a 60\% enhancement
in the magnitude of the Sm $4f$ polarization in the
superlattice compared to bulk SmN, caused by interface
exchange with highly spin-polarized GdN
\cite{Anton_PRB_2013,McNulty_PRB_2015}. This implies that
the bulk Sm $4f$ spin ($m^{4f}_S$) and orbital ($m^{4f}_L$)
moments are not fully polarized ($=5~\mu_B$), but are
significantly reduced from full alignment, consistent with
the $4f$ electrons being dominated by spin-orbit coupling
and crystal field energies. Only two calculations providing
values of the SmN magnetic moment are available, both of
which use density functional theory in the LSDA+$U$
approximation. A report by Larson $\emph{et al.}$
\cite{Larson_PRB_2007} calculated a ferromagnetic ground
state moment nearly equal to the experimental value, with
the opposing $4f$ spin and orbital moments of magnitude
$\approx 4.9~\mu_B$, while the $5d$ and $2p$ states provided
moments of order $0.1~\mu_B$ but with opposite
sign. However, the net moment was found to be parallel to
the spin moment (i.e., spin-dominant). Another study by
Morari \emph{et al.}  \cite{Morari_JPCM_2014} used a variety
of LSDA$+U$ implementations and parameters to find even
larger values for the net moment, though these were also
spin-dominant. The ground state was furthermore
antiferromagnetic, contradicting recent experiments
unambiguously displaying ferromagnetism
\cite{Meyer_PRB_2008,Anton_PRB_2013}.  Other band structure
calculations on EuN \cite{Richter_PRB_2011} and TmN
\cite{Peters_PRB_2014} suggest that dynamical mean-field
theory in the Hubbard-$I$ approximation provided better
agreement with experiment than LSDA$+U$, suggesting
alternative techniques should be explored.

Here we pursue an alternative to band structure calculations
to explain experimental situation in SmN by directly
diagonalizing an effective $4f$ Hamiltonian to yield spin
and orbital moments of the ground state. We neglect
conduction electron contributions which are small for
semiconducting SmN, serving only as a weak correction to our
results. We employ a model of the Sm$^{3+}$ ion in a cubic
crystal field, incorporating excited multiplets following
DeWijn \emph{et al.}  \cite{DeWijn_PSS_1976}, exchange in a
self-consistent mean-field approach \cite{Adachi_PRB_1997},
and spin-orbit coupling. The results demonstrate that the
small orbital-dominant ferromagnetic moment of SmN naturally
emerges from partial-quenching of both orbital and spin $4f$
moments, linked together by the spin-orbit coupling. The
moments were calculated to be less than half of the maximum
$5~\mu_B$ polarization of the $LS$-coupled state. We also
demonstrate a correspondence between the $5d$ polarization
measured by XMCD and the mean-field exchange constant, and
demonstrate that the enhanced $4f$ and $5d$ XMCD
polarization observed in a SmN/GdN superlattice compared to
bulk SmN can be understood within our model. These results
provide the first description of the $4f$ magnetic state of
SmN in quantitative agreement with the various experimental
results and provide crucial insight into recent experiments
showing unconventional twisted magnetization phases and
triplet superconductivity in SmN.

\section{Calculation Procedure}

The magnetic properties of SmN are primarily dictated by the
Sm$^{3+}$ ion, with its atomic-like $4f$ orbitals. Examples
of of Sm$^{3+}$ compounds studied in the past include
metallic Laves-phase compounds such as SmAl$_2$,
Sm$_{1-x}$Gd$_x$Al$_2$, and Sm$_{1-x}$Nd$_x$Al$_2$
\cite{Diepen_PRB_1973,Buschow_PRB_1973,DeWijn_PSS_1976,
  Adachi_Nature_1999,Adachi_PRL_2001,Qiao_PRB_2004,Dhesi_PRB_2010},
CsCl-type structure SmZn and SmCd \cite{Adachi_PRB_1999},
and Sm metal \cite{Adachi_PRB_1997}. These studies all
demonstrate that the magnetic properties of trivalent Sm
compounds are strongly influenced by the first excited
$J=7/2$ multiplet in addition to the ground $J=5/2$
multiplet. There are two reasons for this: 1) the exchange
and Zeeman terms along with the crystal field of the
surrounding anions cause admixtures of the ground and
excited states, and 2) the small ground state Land\'{e}
$g_J$-factor of $\frac{2}{7}$ for Sm$^{3+}$ means the matrix
elements within the $J=5/2$ multiplet are small compared to
$\Delta J = 1$ matrix elements. An important consequence is
the zero-moment ferromagnetism observed in Gd doped SmAl$_2$
\cite{Adachi_Nature_1999, Dhesi_PRB_2010}. In that compound,
the $J$-mixing of the ground and first excited multiplets
causes different temperature dependencies of $m^{4f}_S$ and
$m^{4f}_L$, resulting in a transition from orbital- to
spin-dominant magnetism at a compensation temperature where
the net moment is zero, while long-range ferromagnetic order
is still present.

Crystal field calculations of magnetic properties in
rare-earth systems are usually described within the Steven's
formalism of operator equivalents \cite{Stevens_PPSA_1952,
  Hutchings_SSP_1964}, however these techniques are
insufficient for Sm$^{3+}$ as they only consider the ground
state $J$-multiplet. Here we use an extension of the Steven's
formalism which allows for the inclusion of higher multiplets
\cite{DeWijn_PSS_1976}, derived with the algebraic methods
of tensor operators.

The effective $4f$ Hamiltonian from which we can calculate
the ground-state expectation values includes the spin-orbit
coupling ($\Ham_{SO}$), exchange interaction ($\Ham_{ex}$),
and crystal field ($\Ham_{CF}$). We ignore the Zeeman
coupling for calculation of the spontaneous
moment. Including the $4f$ inter-ion exchange in a
self-consistent mean-field approach
\cite{Adachi_PRB_1997,Adachi_PRB_1999}, the Hamiltonian is
\begin{equation}\label{ex-ham}
\Ham = \Lambda \vec{L}\cdot\vec{S} -\J\ave{S_z}S_z +\Ham_{CF},
\end{equation}
where the spin-orbit coupling constant $\Lambda$ is
approximately $430$~K, calculated from a knowledge of the
energy splittings between $J$-multiplets, expressed as
$\Delta E=E_{J+1}-E_J = \Lambda (J+1)\approx1500~\text{K}$
\cite{Adachi_PRB_1999,Meyer_PRB_2008}. The second term in
Eq. \ref{ex-ham} accounts for exchange, where $\J$ is the
effective exchange constant and $\ave{S_z}$ is the
expectation value of the spin operator. The exchange
constant can be estimated from the paramagnetic Curie
temperature $\Theta$, which in the mean-field approximation
is given by
\begin{equation}
\Theta =
2\J(g_J-1)^2J(J+1)/3k_B.
\end{equation}

The de Gennes factor $G=(g_J-1)^2J(J+1)$ of the ground state
multiplet linking the exchange constant and $\Theta$ is
usually found to describe the trend in $\Theta$ versus $G$
in iso-structural rare-earth compounds. However, this
relation ignores the influence of $J$-mixing by the exchange
field, and thus slightly overestimates $\J$. By considering
the van Vleck temperature-independent contribution to the
susceptibility, one can derive [see Appendix] the
first-order correction to the calculation of $\J$ as
\begin{equation}
  \J= \frac{3}{2} \frac{k_B
    \Theta}{(g_J-1)^2J(J+1)}\left[\frac{1}{1+12k_B\Theta/\Delta
      E}\right].
\end{equation}
Among the rare-earths this correction factor is only
significant for Sm and Eu ions, and in the case of Sm$^{3+}$
it gives a correction factor of $[1+12k_B\Theta/\Delta
E]^{-1}=0.83$, yielding $\J = 7.9~$K. While the correction
here does not qualitatively alter the results of the
following calculations, we include it to demonstrate the
influence of $J$-mixing on the mean-field exchange constant.

The rock-salt structure of SmN results in cubic octahedral
coordination of the Sm$^{3+}$ ion, and we choose the $z$
axis to be along the [111] direction \footnote{We have found
  $z$ parallel to $[111]$ to be the lowest energy
  orientation compared to $[001]$ and $[110]$
  directions. This was also reported in
  \cite{Buschow_PRB_1973} and \cite{Adachi_PRB_1999} for
  Sm$^{3+}$}. The crystal field Hamiltonian is then given by
\cite{Hutchings_SSP_1964,DeWijn_PSS_1976}
\begin{align}\label{H_cf_b}
\begin{aligned}
\Ham_{CF} = 
&-\frac{2}{3}A_4 \sum_i \left( f_{40}(\vec{r}_i) - 20
\sqrt{2} f_{43}(\vec{r}_i)\right)\\ 
&+ \frac{16}{9}A_6\sum_i
\left(f_{60}(\vec{r}_i) + \frac{35\sqrt{2}}{4}f_{63}(\vec{r}_i) +
\frac{77}{8}f_{66}(\vec{r}_i)\right),
\end{aligned}
\end{align}
where the tesseral harmonics $f_{kq}(\vec{r}_i)$ are
renormalized and purely real combinations of spherical
harmonics \cite{Hutchings_SSP_1964}. The parameters $A_4$
and $A_6$ are the fourth- and sixth-order crystal field
parameters discussed below, while the sum is over the five
$4f$ electrons. An arbitrary matrix element of $\Ham_{CF}$
has the form
\begin{equation}
\bra{J'M_{J'}}\Ham_{CF}\ket{JM_J} = \chi_4A_4\ave{r^4}+
\chi_6A_6\ave{r^6},
\end{equation}
where $\chi_k$ contain the angular part of the matrix
elements, described in Ref.  \cite{DeWijn_PSS_1976} for all
the trivalent rare-earth ions. We are then left to determine
$A_4\ave{r^4}$ and $A_6\ave{r^6}$, which can be either treated as
adjustable parameters or calculated within, e.g., the
point-charge model.

After diagonalization of equation \eqref{ex-ham}, the
magnetic moment in the ground state $(T=0)$ is defined by
the expectation value 
\begin{equation}\label{moment_def}
m=\mu_B\ave{L_z+2S_z},
\end{equation}
where $\ave{\cdots}$ is taken within the ground-state,
$\ket{\psi}_{GS} = \sum_J
\sum_{M_J}c_{J,M_J}\ket{JM_J}$. Similarly we have $m^{4f}_L
= \mu_B\ave{L_z}$ and $m^{4f}_S = 2\mu_B\ave{S_z}$. We note
that in Eq. \eqref{moment_def} and for the individual
moments we use the opposite sign convention to define the
magnetic moment compared with the standard definition. If we
restrict to the $J=5/2$ multiplet, the matrix elements are
given by $\ave{JM_J|L_z+2S_z|JM_J} = g_JJ$, with $\ave{L_z}
= (2-g_J)J$ and $\ave{S_z} =(g_J-1)J$. If we consider
instead the lowest three $J=5/2,\;7/2,\;9/2$ multiplets, the
Hamiltonian has dimension $24\times 24$, and the matrix
elements of $L_z,S_z$ and $L_z+2S_z$ can be calculated in
terms of Wigner $3j$ and $6j$ symbols, as given in
Ref. \cite{DeWijn_PSS_1976}. When the exchange term is
included, an initial guess of $\ave{S_z}$ is made, $\Ham$ is
diagonalized, and the resulting ground-state expectation
value $\ave{S_z}$ is then reinserted directly into
Eq. \eqref{ex-ham} repeatedly until convergence is reached,
typically requiring $N\sim 10$ or less iterations.

The effect of including higher multiplets can be seen in
Table 1, where only spin-orbit coupling and exchange are
considered. The free-ion moments ($\J=0$) yield orbital to
spin ratio of $m^{4f}_L/m^{4f}_S = -1.2$, while the
inclusion of the ferromagnetic exchange term with $\J =
7.9$~K reduces the ratio to $m^{4f}_L/m^{4f}_S = -1.175$,
due to the off-diagonal terms in $\Ham_{ex}$ mixing in the
excited multiplet (inclusion of the $J=9/2$ multiplet
returns the same expectation values to the given
precision). The small reduction in the net moment due to the
weak exchange interaction in SmN is clearly unable to
account for the experimental moment of 0.035~$\mu_B$.

We note that if $\Ham_{SO}$ is treated as a small
perturbation to $\Ham_{ex}$, then the net moment is reduced
to zero when higher multiplets are included because $L_z$
and $S_z$ take on fixed values of $M_L$ and $M_S$ (e.g.,
$L_z\ket{\psi} = M_L\ket{\psi}$) for which the ground state
has $M_L=5$ and $M_S=-5/2$, yielding zero net moment. These
values of $m^{4f}_S$ and $m^{4f}_L$ are close to what is
found in the LSDA+$U$ calculations of
Ref. \cite{Larson_PRB_2007}, with $m^{4f}_L/m^{4f}_S =
-4.85/4.91 = -0.98$. In addition,
Ref. \cite{Morari_JPCM_2014}, using various implementations
of LSDA$+U$ and a range of parameters, calculated SmN values
in the range of $m^{4f}_L/m^{4f}_S = -4.57/5.00=-0.85$ to
$m^{4f}_L/m^{4f}_S = -3.46/4.09 = -0.91$. All of
these LSDA$+U$ calculations show a spin-dominant SmN moment,
in contrast to the orbital dominant moment of both the free
ion and SmN.

\begin{table}
  \centering
  \begin{tabular}{lcccc}
    \hline
    $J$-multiplets & $\J$ & $\ave{S_z}$ & $\ave{L_z}$ & $\ave{L_z+2S_z}$ \\
     \hline
     \hline
     $\frac{5}{2}$ & 0& -1.786 & 4.286 & 0.714 \\
     $\frac{5}{2},\frac{7}{2}$ & 0& -1.786 & 4.286 & 0.714 \\
     $\frac{5}{2},\frac{7}{2}$ & 7.9~K& -1.852 & 4.352 & 0.648 \\
     %$\frac{5}{2},\frac{7}{2},\frac{9}{2}$& -1.852 & 4.352 & 0.648 \\
    \hline
  \end{tabular}
  \caption{\label{table_1}Expectation values of the angular momentum operators for the  
    Sm$^{3+}$ ion based on number of $J$-multiplets
    included, only considering the spin-orbit coupling and
    exchange (i.e. omitting the crystal field).}
\label{table1
}
\end{table}

\section{Results}
\begin{figure}
\center\includegraphics[width=0.85\columnwidth]{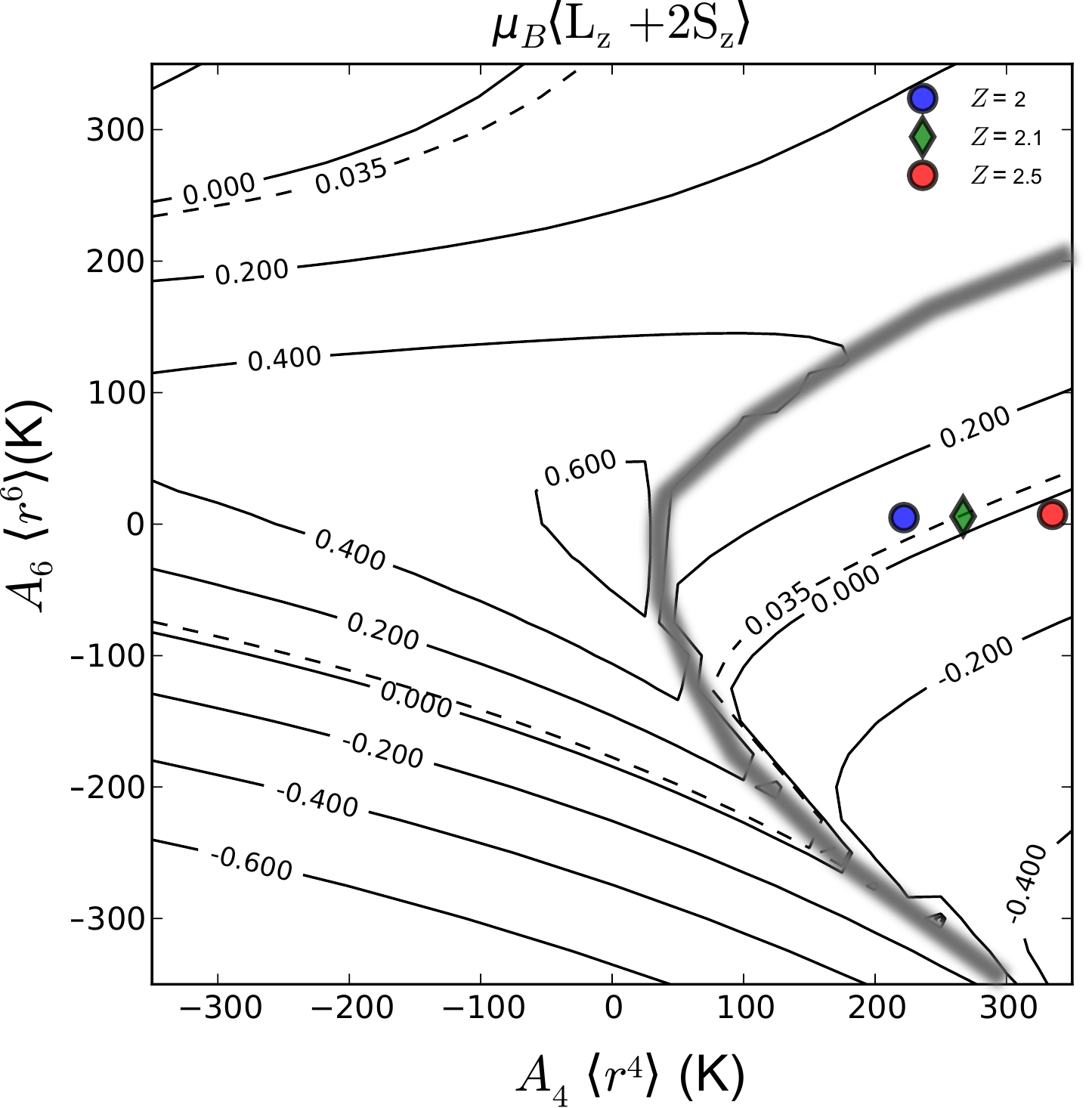}
\caption{\label{moment_plot} Contours of
  $\mu_B\ave{L_z+2S_z}$ for the fourth and sixth order
  crystal field parameters ($\ave{L_z}$ is taken as
  positive). The dashed contour corresponds to
  $\mu_B\ave{L_z+2S_z}= 0.035~\mu_B$, equal to the
  experimental net moment. The heavy grey line demarcates
  the the boundary of the $\Gamma_7$ doublet and $\Gamma_8$
  quartet, the area to the right of this line has a
  $\Gamma_7$ ground state. The symbols represent the net
  $4f$ moment returned using crystal field parameters
  calculated within the point-charge model for three values
  of $Z$.}
\end{figure}

Figure \ref{moment_plot} shows the resulting ground-state
expectation value $\mu_B\ave{L_z+2S_z}$, plotted for crystal
field parameters $A_4\ave{r^4}, A_6\ave{r^6}
\in [-350~\text{K},+350~\text{K}]$ with contours of constant
$\mu_B\ave{L_z + 2S_z}$ plotted, including a dashed contour
corresponding to a calculated net moment equal to the
experimental moment of $0.035~\mu_B$. The ground state
moment was found by diagonalizing the Hamiltonian in
Eq.~\eqref{ex-ham} within the three lowest multiplets
($J=5/2,7/2,9/2$), with $\J = 7.9~$K. To limit the range of
crystal field parameters producing agreement with the
experimental data in Fig. \ref{moment_plot} we must impose
some restrictions. One restriction is provided by
experimental results revealing an effective paramagnetic
moment in SmN close to that calculated from a $\Gamma_7$
ground-state doublet \cite{Meyer_PRB_2008}. The $\Gamma_7$
ground state corresponds to the region to the right of the
heavy line demarcating where the $\Gamma_8$ quartet and
$\Gamma_7$ doublet are degenerate, restricting the possible
crystal field parameters to within this area.

A further refinement of the crystal field parameters can be
made using the the electric-point charge approximation,
which allows us to parametrize the crystal field strength in
terms of a single effective charge $Z$. While this is a
phenomenological approach, it has been shown to provide
reasonable agreement with the variation in crystal field
strength across the light rare-earth pnictides (P, As, Sb,
Bi) \cite{Birgeneau_PRB_1973} as well as some of the other
rare-earth nitrides \cite{Hulliger_HPCRE_1979}, and can thus
provide valuable insight where more sophisticated techniques
have failed. In the point charge model, the coefficients
$A_4$ and $ A_6$ are given by the following expressions in
the case of an octahedrally coordinated rare-earth ion
\cite{Hutchings_SSP_1964,Lea_JPCS_1962},
\begin{equation}
A_4 = \frac{7}{16}\frac{Ze^2}{R^5},\qquad
A_6 = \frac{3}{64}\frac{Ze^2}{R^7}, 
\end{equation}
where $-Z|e|$ is the effective charge of each N ligand, and
$R$ is the separation between the Sm ion and N, given by
half of the experimental SmN lattice-constant
$R=a/2=2.52$~\AA\; \cite{Natali_PMS_2013}. The radial
integrals $\ave{r^k}$ have been calculated in various
approaches, notably non-relativistic Hartree-Fock values
\cite{Freeman_PR_1962} and relativistic Dirac-Fock
calculations \cite{Freeman_JMMM_1979}. The latter of these
yield for Sm$^{3+}$ the values $\ave{r^4} = 2.26$~$a_0^6$ and
$\ave{r^6} = 10.55$~$a_0^6$, where $a_0 = 0.529$~\AA{} is
the Bohr radius \footnote{To less than 1\% error one may
  scale both $\ave{r^4}$ and $\ave{r^6}$ by 1.2 to go from
  Hartree-Fock to Dirac-Fock values, which is equivalent to
  rescaling $Z$. One must therefore scale $Z$ accordingly
  when comparing to values from the literature.}.  We are
then left to determine the effective charge $Z$ of the N
anions. For example, a value of $Z=1.6$ was found to explain
the trend in the fourth order crystal field parameter in the
light rare-earth pnictides (scaled to Dirac-Fock values)
\cite{Birgeneau_PRB_1973}, while larger values of $Z$ are
common in a few of the other rare-earth nitrides, ranging
from 2.3 to 3.6 \cite{Hulliger_HPCRE_1979}. This difference
between the rare-earth nitrides and the pnictides originates
in the greater electronegativity of nitrogen (and hence more
ionic behavior), resulting in the semiconducting nature of
the nitrides, while the other pnictides are metallic.

\begin{table}
  \centering
  $J=5/2,7/2,9/2$ included\\
  \begin{tabular}{ccccccc}
    \hline
    $\J$ (K) & $Z$ & $A_4$ (K) & $A_6$ (K) & $\ave{S_z}$ &
    $\ave{L_z}$& $\ave{L_z+2S_z}$ \\
    \hline
    \hline
    7.9 & 2.1  & 268.9& 5.9 & $-0.976$ & 1.988 & 0.036 \\
    $\J_{SL}$ = 24.5\, & 2.1 &  268.9 & 5.9 &$-1.48$ & 3.11 & 0.16 \\
    \hline\\
\end{tabular}\\
    $J = 5/2$ only  \\
    \begin{tabular}{llccccc}
    \hline
$\J$ (K) & $Z$ & $A_4$ (K) & $A_6$ (K) & $\ave{S_z}$ & $\ave{L_z}$& $\ave{L_z+2S_z}$ \\
    \hline
    \hline
    7.9 & 2.1 & 268.9& 5.9 & $-0.70$ & 1.69 & 0.28 \\
    \hline
    \end{tabular}

    \caption{\label{smn_table}(Top) Expectation values of magnetic moments in 
      SmN with crystal field parameters calculated in the point charge model. (Bottom)
      Expectation values only considering the ground $J=5/2$
      multiplet.}
\end{table}

\begin{figure*}
\center\includegraphics[width=\textwidth]{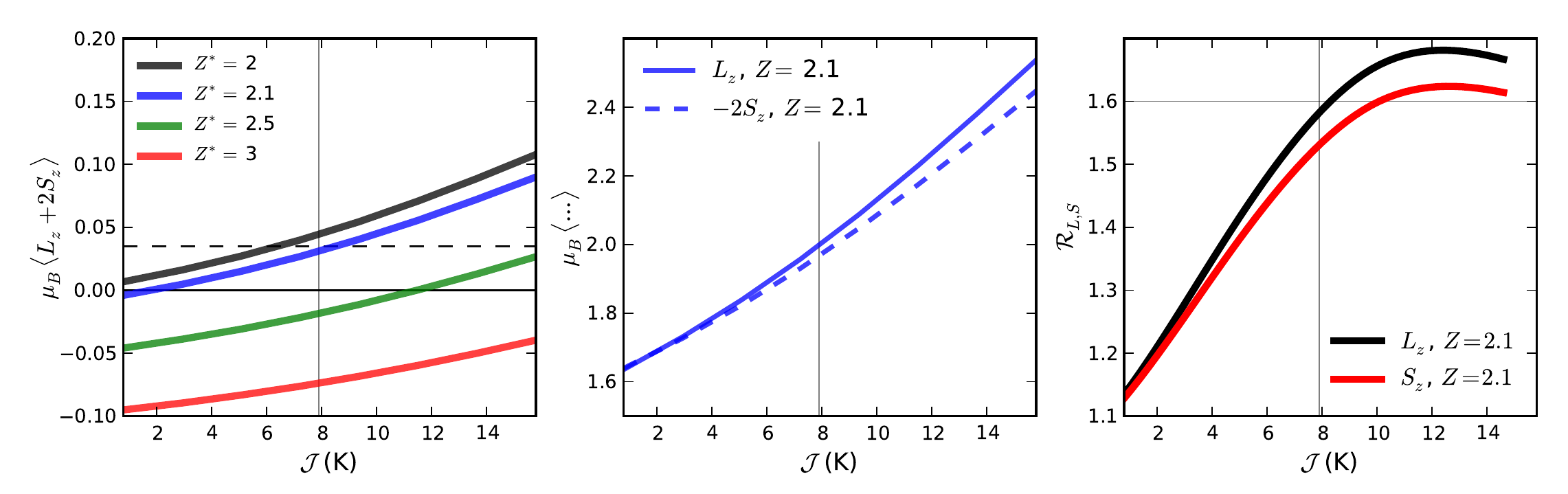}
\caption{\label{point_charge} (a) The ground state $4f$
  expectation value $\ave{L_z+2S_z}$ calculated within the
  lowest three $J$-multiplets for various values of $Z$,
  with $\ave{L_z}$ taken as positive. (b) Expectation values
  $\ave{L_z}$ and $-2\ave{S_z}$ for $Z=2.1$. (c) The
  parameters $\mathcal{R}_L =
  \ave{L_z}_{\J_{SL}}/\ave{L_z}_{\J}$ and$\mathcal{R}_S =
  \ave{S_z}_{\J_{SL}}/\ave{S_z}_{\J}$ representing the
  enhancement in the superlattice.}
\end{figure*}

In Figure \ref{point_charge} we present ground-state
expectation values calculated using the point charge model
with various values of $Z$ to parametrize the crystal field
strength. The expectation values are plotted against $\J$,
centered around the experimentally-derived value of $\J =
7.9~$K, in order to show the influence of exchange.  The
first panel of Fig.~\ref{point_charge} shows that the
experimental moment of $0.035~\mu_B$ is reproduced by using
parameters obtained from the point-charge model with $Z =
2.1$ and $\J=7.9$~K (see also Table \ref{smn_table}). The
value of $Z=2.1$ fits in with the trends observed in the
rare-earth nitrides and pnictides, larger than the $Z=1.6$
value found for the light rare-earth pnictides and slightly
less than the $Z=2.3$ - $3.6$ range found in a number of
rare-earth nitrides, as noted above.  This result
demonstrates that using reasonable parameters, the model
reproduces the sign and magnitude of the experimental net
moment of SmN. Furthermore, in Figure \ref{point_charge} (b)
we see that the spin and orbital moments have magnitudes of
$\approx 2~\mu_B$, strongly quenched compared to both the
free-ion moments and the values given in
Refs. \cite{Larson_PRB_2007,Morari_JPCM_2014}

Figure \ref{point_charge} (b) also shows that for $Z=2.1$
and $\J = 7.9$~K the ratio $m^{4f}_L/m^{4f}_S = -1.02$ is
suppressed from the free-ion value of $-1.2$. This
suppressed ratio only occurs when there is $J$-mixing. For
example, if we are restricted to the ground-state
$J$-multiplet, the Wigner-Eckart theorem assures that the
ratio $\ave{L_z}/\ave{2S_z}= (2-g_J)/[2(g_J-1)] = -1.2$ is
fixed and independent of the parameters in the
Hamiltonian. From this point of view we can see that the net
moment is strongly influenced not only by the overall
reduction of the spin and orbital moments, but also by their
changing ratio due to $J$-mixing.

While the net $4f$ moment calculated above can be compared
to the experimental net moment (ignoring other
contributions), it is more difficult to obtain experimental
values of $m^{4f}_S$ and $m^{4f}_L$. The ideal method is
through x-ray magnetic circular dichroism (XMCD) at the
rare-earth M-edge, which is in principle able to determine
$m^{4f}_S$ and $m^{4f}_L$ experimentally via the XMCD sum
rules for electric-dipole
transitions \cite{Thole_PRL_1992,Carra_PRL_1993}. Experimental
limitations have so far prevented a quantitative sum rule
analysis in SmN, but we may still extract meaningful
conclusions by exploring the Sm L$_3$-edge XMCD measured in
a SmN film and a SmN/GdN superlattice, reported in
Ref. \cite{McNulty_PRB_2015}.

\begin{figure}
\center\includegraphics[width=\columnwidth]{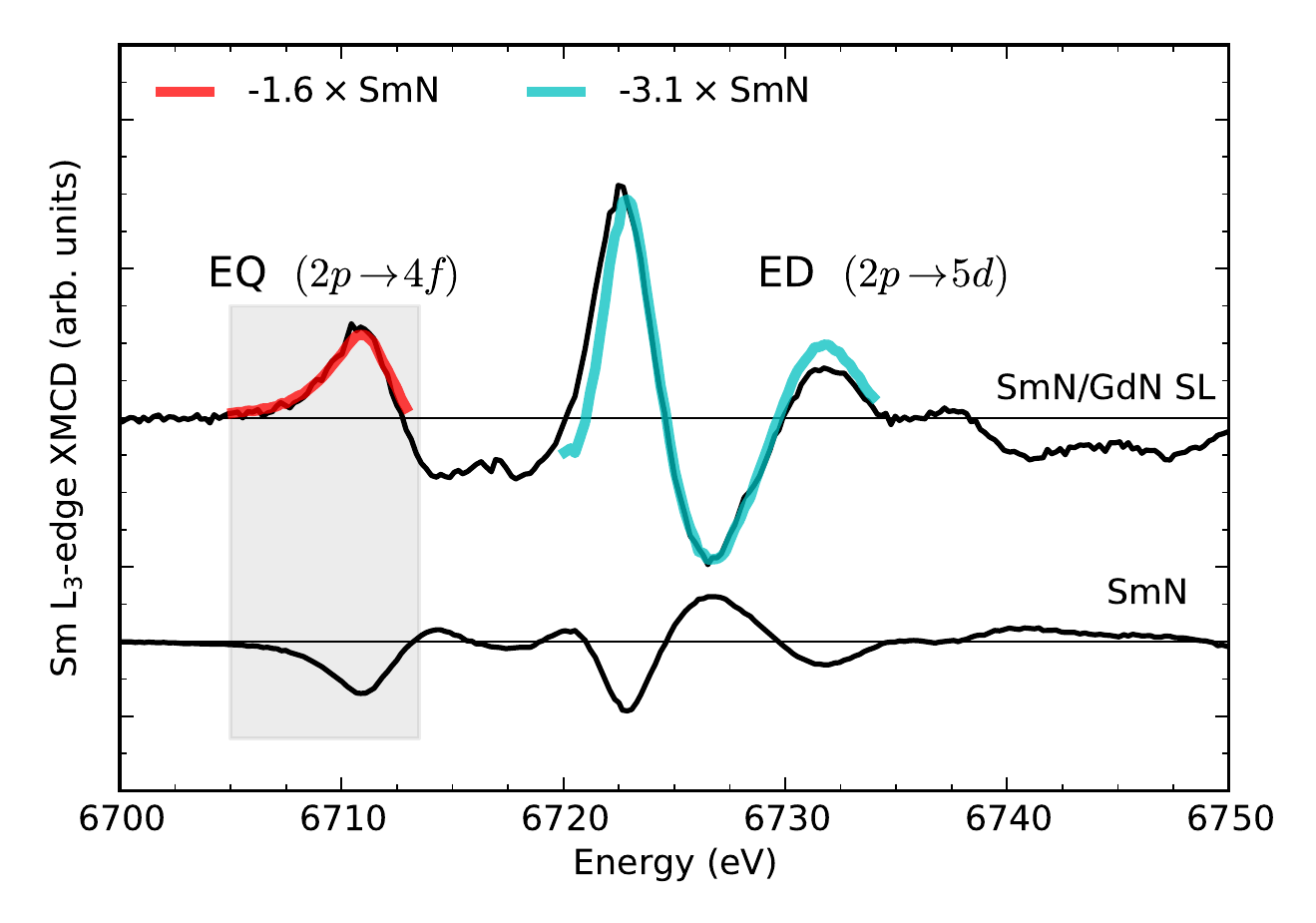}
\caption{\label{xmcd} Sm L$_3$ edge XMCD data from
  Ref. \cite{McNulty_PRB_2015}, showing spectra taken at
  15~K and 6~T for SmN and a SmN/GdN superlattice. The
  scaling of EQ and ED signals between
  samples is also displayed.}
\end{figure}

Figure \ref{xmcd} shows the Sm L$_3$ edge XMCD data taken
from Ref. \cite{McNulty_PRB_2015}, which showing the spectra from
a bulk SmN sample and a 12$\times$[1.5 nm SmN/9 nm GdN]
superlattice. The Sm L$_3$ edge is dominated by
electric-dipole (ED) $(2p\rightarrow 5d)$ transitions at
high energy. These ED transitions provide information on the
sign and magnitude of the polarization of the unoccupied Sm
$5d$ states which mediate exchange between the localized
$4f$ moments. At lower energy, the electric-quadrupole (EQ)
transitions ($2p\rightarrow 4f$) are visible, providing sign
and magnitude information on the $4f$ shell alignment. To a
good approximation, the integral over the Sm L$_3$ EQ
spectrum scales linearly with $\ave{S_z}$ and $\ave{L_z}$,
which can be seen from the XMCD sum rules derived for EQ
transitions \cite{Carra_PB_1993}. As discussed in Refs
\cite{Anton_PRB_2013,McNulty_PRB_2015}, the sign change
between spectra indicates the magnetization of the thin SmN
layers are forced into alignment with the GdN by strong
interface-exchange in the superlattice. In addition to the
sign change, the ED signal is enhanced by a factor of 3.1 in
the superlattice, while the EQ feature is only enhanced by a
factor of 1.6.

First, we point out that an enhancement of the EQ signal in
the superlattice SmN layers indicates that the $4f$ spin and
orbital moments cannot be maximal ($=5~\mu_B$) in the bulk
SmN, as this would preclude any enhancement.  If we assume
the superlattice SmN has spin/orbital moments maximized at
$5~\mu_B$, an upper bound on the bulk SmN spin/orbital
moments can be determined as
$|m^{4f,\text{max}}_{L,S}|\approx 5~\mu_B/1.6 = 3.1
~\mu_B$. This upper bound is already well \emph{below} the
LSDA$+U$ results of
Refs. \cite{Larson_PRB_2007,Morari_JPCM_2014}, but is still
larger than the present calculations which show moments of
magnitude $\approx 2~\mu_B$. Clearly, even within the
superlattice, the SmN spin and orbital moments are not fully
polarized.

A more quantitative analysis of the observed ED and EQ
enhancement factors can also be made.  The strong intra-ion
$4f$-$5d$ exchange mechanism of the rare-earths means that
inter-ion $4f$-$4f$ exchange proceeds via the $5d$ states,
resulting in an effective $4f$ inter-ion exchange constant
($\J)$ proportional to the Sm $5d$ polarization given by the
ED XMCD. From this identification it is clear that the
average superlattice SmN exchange constant $\J_{SL}$ is
enhanced by a factor of 3.1 (the ED enhancement factor) over
the bulk SmN exchange $\J$. We can then calculate the
expectation values for the superlattice SmN using $\J_{SL} =
3.1 \times\J$ as the exchange constant. The enhancement of
the orbital and spin moments in the superlattice compared to
bulk SmN is then determined by the ratios
$\mathcal{R}_L=\ave{L_z}_{\J_{SL}}/\ave{L_z}_{\J}$ and
$\mathcal{R}_S=\ave{S_z}_{\J_{SL}}/\ave{S_z}_{\J}$.  The
ratios are plotted as a function of $\J$ in the third panel
of Figure \ref{point_charge}. The results show that for
$\J=7.9$~K and $\J_{SL} = 24.5$~K (see also Table
\ref{smn_table}), values of $\mathcal{R}_L=1.56$ and
$\mathcal{R}_S=1.52$ are returned, closely reproducing the
factor of 1.6 enhancement in the EQ signal. As noted above,
$\ave{S_z}$ and $\ave{L_z}$ scale nearly linearly with the
EQ spectrum, demonstrating that the model describes the
values $4f$ moments quite well.

% The small
% differences in the values of $\mathcal{R}_L$ and
% $\mathcal{R}_S$ are due
% to $J$-mixing; restricting to the ground $J=5/2$-multiplet
% the ratios are $\mathcal{R}_L=\mathcal{R}_S=1.53$. 

\section{Discussion}

The ability of the model to reproduce not only the net
experimental moment in bulk SmN (sign and magnitude), but
also the change in $\ave{S_z}$ and $\ave{L_z}$ due to
interface exchange in the superlattice shows it captures the
essential features of the SmN ground state. The results
further show that contributions to the net moment from
conduction electrons and other sources provide only small
corrections to our results. The moderate quenching of the
spin and orbital moments to magnitudes of $\sim 2~\mu_B$ in
bulk SmN, enhanced to $\sim 3~\mu_B$ in SmN layers within
the SmN/GdN superlattice are significantly less than the
values given by LSDA$+U$ calculations. However, this
quenching in SmN fits in with recent experiments
\cite{Cortie_PRB_2014} on HoN, ErN and DyN (all heavy
rare-earth and spin-dominant systems) which also display
significant reduction of the net moment compared to the
Hund's rules values and the LSDA$+U$ calculations in Ref
\cite{Larson_PRB_2007}. There the authors suggest the small
ordered moments can be explained by full quenching of
$\ave{L_z}$, but this seems unlikely due to the large
spin-orbit coupling energy of the $4f$ shell. It is more
likely that in these RENs partial quenching of both
$\ave{S_z}$ and $\ave{L_z}$ takes place, as we have shown
for SmN. The quenching of moments in rare-earth systems with
small exchange energies due to crystal fields has been known
for some time \cite{Trammell_PR_1963}, suggesting that
density-functional theory calculations in the RENs should be
revisited to obtain more accurate descriptions of the
magnetic structure.

The values of the spin and orbital moments of $\approx
2~\mu_B$ in bulk SmN and moments of $\approx 3~\mu_B$ in the
superlattice SmN layers are also relevant to the the recent
observation of superconductivity coinciding with
ferromagnetic order in SmN
\cite{Anton_Arxiv_2015}. Superconductivity at $T_c\approx
3$~K was observed in heavily doped SmN film, while in a
$12\times (10 \text{ nm GdN}/5 \text{ nm SmN})$ superlattice
a superconducting transition occurred at $T_c \approx 5$~K,
with a much larger critical field. The heavy doping is
expected to reduce the net moment (the conduction electron
contribution $m_{\sigma}$ is parallel to $m^{4f}_S$), and
could even lead to a spin-dominant moment at high enough
doping levels. The more robust superconductivity (which was
shown to involve $4f$ states) in the superlattice is likely
due to enhancement of the $4f$ spin and orbital moments due
interface exchange with GdN. The enhanced $4f$ moments may
then be more robust against disorder, leading to a higher
$T_c$ and a larger critical field. We note that the observed
$4f$ enhancement from XMCD data of
Ref. \cite{McNulty_PRB_2015} given in Fig. \ref{xmcd} is
from SmN 1.5~nm thick, while the superconducting SmN layers
in the SmN/GdN superlattice of Ref. \cite{Anton_Arxiv_2015}
were 5~nm thick, indicating that the effect of interface
exchange is smaller (averaged over the SmN layers) in the
superconducting samples, so that the average enhancement in
$4f$ polarization is probably less than 60\% enhancement
observed in the XMCD data.

Also closely related to the current calculations is the
twisted magnetization phase observed in a SmN/GdN bilayer
\cite{McNulty_PRB_2015}, in which the SmN is exchange
coupled to GdN and one interface, and results in a rotating
magnetization as the SmN moments experience exchange-Zeeman
competition. This twisted magnetization phase may be viewed
as a type of unconventional exchange-spring; it is the
magnetically \emph{hard} SmN layer which develops the spring
(or twist), while the soft GdN layer is fixed. The SmN/GdN
system is also semiconducting, in contrast to conventional
exchange spring-systems which are metallic. These features
provide an opportunity for tunneling magnetoresistance
experiments which may show unusual behavior. The current
calculations revealing quenched moments will thus provide an
important aide in the quantitative interpretation of
tunneling magnetoresistance experiments involving SmN/GdN
heterostructures, and should allow for more accurate
micromagnetic simulations of the twisted magnetization
phase.

\section{Conclusion}

We have shown that the small orbital-dominant SmN moment can
be explained by consideration of an effective $4f$
Hamiltonian incorporating excited $J$-multiplets, exchange
in a self-consistent mean-field approach, spin-orbit
coupling, and crystal field terms with parameters estimated
from the point-charge model. The magnitudes of the spin and
orbital moments were shown to be partially quenched to
$\approx 2~\mu_B$ in bulk SmN. By linking the Sm $5d$ XMCD
polarization to the effective exchange constant, we showed
that our model reproduces the enhancement in the Sm $4f$
electric-quadrupole XMCD signal in the SmN/GdN superlattice
compared to bulk SmN, due to the enhanced spin and orbital
moments of magnitude $\approx 3~\mu_B$. The magnitudes of
these moments are far less than values obtained from
free-ion or LSDA$+U$ calculations, and suggests that
theoretical band structure investigations of SmN and other
rare-earth nitrides should be revisited. The partially
quenched spin and orbital moments in SmN and their
enhancement in the SmN/GdN superlattice corroborate the
previous report that the $4f$ states are crucial for
superconductivity in SmN. Finally, the values of the
magnetic moments presented here will aide in quantitative
interpretation of future tunelling magnetoresistance
experiments involving SmN and SmN/GdN heterostructures.

\section{Appendix}
The magnetic susceptibility $\chi$ may be written in terms
of the susceptibility in the absence of exchange, $\chi_0$,
and the molecular field constant $\lambda$ (in the
mean-field approximation) as
\begin{equation}
\frac{1}{\chi}=\frac{1}{\chi_0} -\lambda =
\frac{1}{\chi_{C} +\chi_{VV}} - \lambda,
\end{equation}
where $\lambda = (V/N)2\J(g_J-1)^2/(g_J\mu_B)^2$. The
Curie susceptibility is 
\begin{equation}
\chi_C = \frac{C}{T}=\frac{N}{V} \frac{g^2_J\mu_B^2 J(J+1)}{3k_BT}
\end{equation}
and the van Vleck susceptibility for Sm$^{3+}$ is 
\begin{equation}
  \chi_{VV} = \frac{N}{V}\frac{20\mu_B^2}{7\Delta E}.
\end{equation}

The presence of $\chi_{VV}$ modifies the usual Curie-Weiss
temperature $\Theta = 2\J(g_J-1)^2J(J+1)/3k_B$, and thus our
estimate of $\J$. To first order, the correction to the
critical temperature $\Theta$ is given by the solution to
$T= (C +\chi_{VV} T)\lambda$. Solving for $\J$ yields
\begin{equation}
  \J= \frac{3}{2} \frac{k_B
    \Theta}{(g_J-1)^2J(J+1)}\left[\frac{1}{1+12k_B\Theta/\Delta
      E}\right].
\end{equation}

% \printbibliography
% %%%%%%%%%%%%%%%%%%%%%%%%%%%%%%%%%%%%%%%%%%%%%%%%%%%%%%%%%%%%%%%%%%%%%%%
% %%	END
% %%%%%%%%%%%%%%%%%%%%%%%%%%%%%%%%%%%%%%%%%%%%%%%%%%%%%%%%%%%%%%%%%%%%%%%
% \end{document}

\section*{Acknowledgements}
We acknowledge financial support from the New Zealand
Foundation for Research, Science and Technology (Grant No.
VICX0808) and the Marsden Fund (Grant No. 08-VUW- 030). The
MacDiarmid Institute is supported by the New Zealand Centres
of Research Excellence Fund.

\bibliography{/home/james/Documents/drafts/xtal_field/library_2015}

\begin{thebibliography}{48}%
\makeatletter
\providecommand \@ifxundefined [1]{%
 \@ifx{#1\undefined}
}%
\providecommand \@ifnum [1]{%
 \ifnum #1\expandafter \@firstoftwo
 \else \expandafter \@secondoftwo
 \fi
}%
\providecommand \@ifx [1]{%
 \ifx #1\expandafter \@firstoftwo
 \else \expandafter \@secondoftwo
 \fi
}%
\providecommand \natexlab [1]{#1}%
\providecommand \enquote  [1]{``#1''}%
\providecommand \bibnamefont  [1]{#1}%
\providecommand \bibfnamefont [1]{#1}%
\providecommand \citenamefont [1]{#1}%
\providecommand \href@noop [0]{\@secondoftwo}%
\providecommand \href [0]{\begingroup \@sanitize@url \@href}%
\providecommand \@href[1]{\@@startlink{#1}\@@href}%
\providecommand \@@href[1]{\endgroup#1\@@endlink}%
\providecommand \@sanitize@url [0]{\catcode `\\12\catcode `\$12\catcode
  `\&12\catcode `\#12\catcode `\^12\catcode `\_12\catcode `\%12\relax}%
\providecommand \@@startlink[1]{}%
\providecommand \@@endlink[0]{}%
\providecommand \url  [0]{\begingroup\@sanitize@url \@url }%
\providecommand \@url [1]{\endgroup\@href {#1}{\urlprefix }}%
\providecommand \urlprefix  [0]{URL }%
\providecommand \Eprint [0]{\href }%
\providecommand \doibase [0]{http://dx.doi.org/}%
\providecommand \selectlanguage [0]{\@gobble}%
\providecommand \bibinfo  [0]{\@secondoftwo}%
\providecommand \bibfield  [0]{\@secondoftwo}%
\providecommand \translation [1]{[#1]}%
\providecommand \BibitemOpen [0]{}%
\providecommand \bibitemStop [0]{}%
\providecommand \bibitemNoStop [0]{.\EOS\space}%
\providecommand \EOS [0]{\spacefactor3000\relax}%
\providecommand \BibitemShut  [1]{\csname bibitem#1\endcsname}%
\let\auto@bib@innerbib\@empty
%</preamble>
\bibitem [{\citenamefont {Natali}\ \emph
  {et~al.}(2013{\natexlab{a}})\citenamefont {Natali}, \citenamefont {Ruck},
  \citenamefont {Plank}, \citenamefont {Trodahl}, \citenamefont {Granville},
  \citenamefont {Meyer},\ and\ \citenamefont {Lambrecht}}]{Natali_PMS_2013}%
  \BibitemOpen
  \bibfield  {author} {\bibinfo {author} {\bibfnamefont {F.}~\bibnamefont
  {Natali}}, \bibinfo {author} {\bibfnamefont {B.~J.}\ \bibnamefont {Ruck}},
  \bibinfo {author} {\bibfnamefont {N.~O.~V.}\ \bibnamefont {Plank}}, \bibinfo
  {author} {\bibfnamefont {H.~J.}\ \bibnamefont {Trodahl}}, \bibinfo {author}
  {\bibfnamefont {S.}~\bibnamefont {Granville}}, \bibinfo {author}
  {\bibfnamefont {C.}~\bibnamefont {Meyer}}, \ and\ \bibinfo {author}
  {\bibfnamefont {W.~R.~L.}\ \bibnamefont {Lambrecht}},\ }\href {\doibase
  http://dx.doi.org/10.1016/j.pmatsci.2013.06.002} {\bibfield  {journal}
  {\bibinfo  {journal} {Prog. Mater Sci.}\ }\textbf {\bibinfo {volume} {58}},\
  \bibinfo {pages} {1316 } (\bibinfo {year} {2013}{\natexlab{a}})}\BibitemShut
  {NoStop}%
\bibitem [{\citenamefont {Leuenberger}\ \emph {et~al.}(2005)\citenamefont
  {Leuenberger}, \citenamefont {Parge}, \citenamefont {Felsch}, \citenamefont
  {Fauth},\ and\ \citenamefont {Hessler}}]{Leuenberger_PRB_2005}%
  \BibitemOpen
  \bibfield  {author} {\bibinfo {author} {\bibfnamefont {F.}~\bibnamefont
  {Leuenberger}}, \bibinfo {author} {\bibfnamefont {A.}~\bibnamefont {Parge}},
  \bibinfo {author} {\bibfnamefont {W.}~\bibnamefont {Felsch}}, \bibinfo
  {author} {\bibfnamefont {K.}~\bibnamefont {Fauth}}, \ and\ \bibinfo {author}
  {\bibfnamefont {M.}~\bibnamefont {Hessler}},\ }\href {\doibase
  10.1103/PhysRevB.72.014427} {\bibfield  {journal} {\bibinfo  {journal} {Phys.
  Rev. B}\ }\textbf {\bibinfo {volume} {72}},\ \bibinfo {pages} {014427}
  (\bibinfo {year} {2005})}\BibitemShut {NoStop}%
\bibitem [{\citenamefont {Granville}\ \emph {et~al.}(2006)\citenamefont
  {Granville}, \citenamefont {Ruck}, \citenamefont {Budde}, \citenamefont
  {Koo}, \citenamefont {Pringle}, \citenamefont {Kuchler}, \citenamefont
  {Preston}, \citenamefont {Housden}, \citenamefont {Lund}, \citenamefont
  {Bittar}, \citenamefont {Williams},\ and\ \citenamefont
  {Trodahl}}]{Granville_PRB_2006}%
  \BibitemOpen
  \bibfield  {author} {\bibinfo {author} {\bibfnamefont {S.}~\bibnamefont
  {Granville}}, \bibinfo {author} {\bibfnamefont {B.~J.}\ \bibnamefont {Ruck}},
  \bibinfo {author} {\bibfnamefont {F.}~\bibnamefont {Budde}}, \bibinfo
  {author} {\bibfnamefont {A.}~\bibnamefont {Koo}}, \bibinfo {author}
  {\bibfnamefont {D.~J.}\ \bibnamefont {Pringle}}, \bibinfo {author}
  {\bibfnamefont {F.}~\bibnamefont {Kuchler}}, \bibinfo {author} {\bibfnamefont
  {A.~R.~H.}\ \bibnamefont {Preston}}, \bibinfo {author} {\bibfnamefont
  {D.~H.}\ \bibnamefont {Housden}}, \bibinfo {author} {\bibfnamefont
  {N.}~\bibnamefont {Lund}}, \bibinfo {author} {\bibfnamefont {A.}~\bibnamefont
  {Bittar}}, \bibinfo {author} {\bibfnamefont {G.~V.~M.}\ \bibnamefont
  {Williams}}, \ and\ \bibinfo {author} {\bibfnamefont {H.~J.}\ \bibnamefont
  {Trodahl}},\ }\href {\doibase 10.1103/PhysRevB.73.235335} {\bibfield
  {journal} {\bibinfo  {journal} {Phys. Rev. B}\ }\textbf {\bibinfo {volume}
  {73}},\ \bibinfo {pages} {235335} (\bibinfo {year} {2006})}\BibitemShut
  {NoStop}%
\bibitem [{\citenamefont {Preston}\ \emph {et~al.}(2007)\citenamefont
  {Preston}, \citenamefont {Granville}, \citenamefont {Housden}, \citenamefont
  {Ludbrook}, \citenamefont {Ruck}, \citenamefont {Trodahl}, \citenamefont
  {Bittar}, \citenamefont {Williams}, \citenamefont {Downes}, \citenamefont
  {DeMasi}, \citenamefont {Zhang}, \citenamefont {Smith},\ and\ \citenamefont
  {Lambrecht}}]{Preston_PRB_2007}%
  \BibitemOpen
  \bibfield  {author} {\bibinfo {author} {\bibfnamefont {A.~R.~H.}\
  \bibnamefont {Preston}}, \bibinfo {author} {\bibfnamefont {S.}~\bibnamefont
  {Granville}}, \bibinfo {author} {\bibfnamefont {D.~H.}\ \bibnamefont
  {Housden}}, \bibinfo {author} {\bibfnamefont {B.}~\bibnamefont {Ludbrook}},
  \bibinfo {author} {\bibfnamefont {B.~J.}\ \bibnamefont {Ruck}}, \bibinfo
  {author} {\bibfnamefont {H.~J.}\ \bibnamefont {Trodahl}}, \bibinfo {author}
  {\bibfnamefont {A.}~\bibnamefont {Bittar}}, \bibinfo {author} {\bibfnamefont
  {G.~V.~M.}\ \bibnamefont {Williams}}, \bibinfo {author} {\bibfnamefont
  {J.~E.}\ \bibnamefont {Downes}}, \bibinfo {author} {\bibfnamefont
  {A.}~\bibnamefont {DeMasi}}, \bibinfo {author} {\bibfnamefont
  {Y.}~\bibnamefont {Zhang}}, \bibinfo {author} {\bibfnamefont {K.~E.}\
  \bibnamefont {Smith}}, \ and\ \bibinfo {author} {\bibfnamefont {W.~R.~L.}\
  \bibnamefont {Lambrecht}},\ }\href {\doibase 10.1103/PhysRevB.76.245120}
  {\bibfield  {journal} {\bibinfo  {journal} {Phys. Rev. B}\ }\textbf {\bibinfo
  {volume} {76}},\ \bibinfo {pages} {245120} (\bibinfo {year}
  {2007})}\BibitemShut {NoStop}%
\bibitem [{\citenamefont {Meyer}\ \emph {et~al.}(2008)\citenamefont {Meyer},
  \citenamefont {Ruck}, \citenamefont {Zhong}, \citenamefont {Granville},
  \citenamefont {Preston}, \citenamefont {Williams},\ and\ \citenamefont
  {Trodahl}}]{Meyer_PRB_2008}%
  \BibitemOpen
  \bibfield  {author} {\bibinfo {author} {\bibfnamefont {C.}~\bibnamefont
  {Meyer}}, \bibinfo {author} {\bibfnamefont {B.~J.}\ \bibnamefont {Ruck}},
  \bibinfo {author} {\bibfnamefont {J.}~\bibnamefont {Zhong}}, \bibinfo
  {author} {\bibfnamefont {S.}~\bibnamefont {Granville}}, \bibinfo {author}
  {\bibfnamefont {A.~R.~H.}\ \bibnamefont {Preston}}, \bibinfo {author}
  {\bibfnamefont {G.~V.~M.}\ \bibnamefont {Williams}}, \ and\ \bibinfo {author}
  {\bibfnamefont {H.~J.}\ \bibnamefont {Trodahl}},\ }\href {\doibase
  10.1103/PhysRevB.78.174406} {\bibfield  {journal} {\bibinfo  {journal} {Phys.
  Rev. B}\ }\textbf {\bibinfo {volume} {78}},\ \bibinfo {pages} {174406}
  (\bibinfo {year} {2008})}\BibitemShut {NoStop}%
\bibitem [{\citenamefont {Meyer}\ \emph {et~al.}(2010)\citenamefont {Meyer},
  \citenamefont {Ruck}, \citenamefont {Preston}, \citenamefont {Granville},
  \citenamefont {Williams},\ and\ \citenamefont {Trodahl}}]{Meyer_JMMM_2010}%
  \BibitemOpen
  \bibfield  {author} {\bibinfo {author} {\bibfnamefont {C.}~\bibnamefont
  {Meyer}}, \bibinfo {author} {\bibfnamefont {B.~J.}\ \bibnamefont {Ruck}},
  \bibinfo {author} {\bibfnamefont {A.~R.~H.}\ \bibnamefont {Preston}},
  \bibinfo {author} {\bibfnamefont {S.}~\bibnamefont {Granville}}, \bibinfo
  {author} {\bibfnamefont {G.~V.~M.}\ \bibnamefont {Williams}}, \ and\ \bibinfo
  {author} {\bibfnamefont {H.~J.}\ \bibnamefont {Trodahl}},\ }\href {\doibase
  http://dx.doi.org/10.1016/j.jmmm.2010.01.016} {\bibfield  {journal} {\bibinfo
   {journal} {J. Magn. Magn. Mater.}\ }\textbf {\bibinfo {volume} {322}},\
  \bibinfo {pages} {1973 } (\bibinfo {year} {2010})}\BibitemShut {NoStop}%
\bibitem [{\citenamefont {Richter}\ \emph {et~al.}(2011)\citenamefont
  {Richter}, \citenamefont {Ruck}, \citenamefont {Simpson}, \citenamefont
  {Natali}, \citenamefont {Plank}, \citenamefont {Azeem}, \citenamefont
  {Trodahl}, \citenamefont {Preston}, \citenamefont {Chen}, \citenamefont
  {McNulty}, \citenamefont {Smith}, \citenamefont {Tadich}, \citenamefont
  {Cowie}, \citenamefont {Svane}, \citenamefont {van Schilfgaarde},\ and\
  \citenamefont {Lambrecht}}]{Richter_PRB_2011}%
  \BibitemOpen
  \bibfield  {author} {\bibinfo {author} {\bibfnamefont {J.~H.}\ \bibnamefont
  {Richter}}, \bibinfo {author} {\bibfnamefont {B.~J.}\ \bibnamefont {Ruck}},
  \bibinfo {author} {\bibfnamefont {M.}~\bibnamefont {Simpson}}, \bibinfo
  {author} {\bibfnamefont {F.}~\bibnamefont {Natali}}, \bibinfo {author}
  {\bibfnamefont {N.~O.~V.}\ \bibnamefont {Plank}}, \bibinfo {author}
  {\bibfnamefont {M.}~\bibnamefont {Azeem}}, \bibinfo {author} {\bibfnamefont
  {H.~J.}\ \bibnamefont {Trodahl}}, \bibinfo {author} {\bibfnamefont
  {A.~R.~H.}\ \bibnamefont {Preston}}, \bibinfo {author} {\bibfnamefont
  {B.}~\bibnamefont {Chen}}, \bibinfo {author} {\bibfnamefont {J.}~\bibnamefont
  {McNulty}}, \bibinfo {author} {\bibfnamefont {K.~E.}\ \bibnamefont {Smith}},
  \bibinfo {author} {\bibfnamefont {A.}~\bibnamefont {Tadich}}, \bibinfo
  {author} {\bibfnamefont {B.}~\bibnamefont {Cowie}}, \bibinfo {author}
  {\bibfnamefont {A.}~\bibnamefont {Svane}}, \bibinfo {author} {\bibfnamefont
  {M.}~\bibnamefont {van Schilfgaarde}}, \ and\ \bibinfo {author}
  {\bibfnamefont {W.~R.~L.}\ \bibnamefont {Lambrecht}},\ }\href {\doibase
  10.1103/PhysRevB.84.235120} {\bibfield  {journal} {\bibinfo  {journal} {Phys.
  Rev. B}\ }\textbf {\bibinfo {volume} {84}},\ \bibinfo {pages} {235120}
  (\bibinfo {year} {2011})}\BibitemShut {NoStop}%
\bibitem [{\citenamefont {Le~Binh}\ \emph {et~al.}(2013)\citenamefont
  {Le~Binh}, \citenamefont {Ruck}, \citenamefont {Natali}, \citenamefont
  {Warring}, \citenamefont {Trodahl}, \citenamefont {Anton}, \citenamefont
  {Meyer}, \citenamefont {Ranno}, \citenamefont {Wilhelm},\ and\ \citenamefont
  {Rogalev}}]{Binh_PRL_2013}%
  \BibitemOpen
  \bibfield  {author} {\bibinfo {author} {\bibfnamefont {D.}~\bibnamefont
  {Le~Binh}}, \bibinfo {author} {\bibfnamefont {B.~J.}\ \bibnamefont {Ruck}},
  \bibinfo {author} {\bibfnamefont {F.}~\bibnamefont {Natali}}, \bibinfo
  {author} {\bibfnamefont {H.}~\bibnamefont {Warring}}, \bibinfo {author}
  {\bibfnamefont {H.~J.}\ \bibnamefont {Trodahl}}, \bibinfo {author}
  {\bibfnamefont {E.-M.}\ \bibnamefont {Anton}}, \bibinfo {author}
  {\bibfnamefont {C.}~\bibnamefont {Meyer}}, \bibinfo {author} {\bibfnamefont
  {L.}~\bibnamefont {Ranno}}, \bibinfo {author} {\bibfnamefont
  {F.}~\bibnamefont {Wilhelm}}, \ and\ \bibinfo {author} {\bibfnamefont
  {A.}~\bibnamefont {Rogalev}},\ }\href {\doibase
  10.1103/PhysRevLett.111.167206} {\bibfield  {journal} {\bibinfo  {journal}
  {Phys. Rev. Lett.}\ }\textbf {\bibinfo {volume} {111}},\ \bibinfo {pages}
  {167206} (\bibinfo {year} {2013})}\BibitemShut {NoStop}%
\bibitem [{\citenamefont {Azeem}\ \emph {et~al.}(2013)\citenamefont {Azeem},
  \citenamefont {Ruck}, \citenamefont {Le~Binh}, \citenamefont {Warring},
  \citenamefont {Trodahl}, \citenamefont {Strickland}, \citenamefont {Koo},
  \citenamefont {Goian},\ and\ \citenamefont {Kamba}}]{Azeem_JAP_2013}%
  \BibitemOpen
  \bibfield  {author} {\bibinfo {author} {\bibfnamefont {M.}~\bibnamefont
  {Azeem}}, \bibinfo {author} {\bibfnamefont {B.~J.}\ \bibnamefont {Ruck}},
  \bibinfo {author} {\bibfnamefont {D.}~\bibnamefont {Le~Binh}}, \bibinfo
  {author} {\bibfnamefont {H.}~\bibnamefont {Warring}}, \bibinfo {author}
  {\bibfnamefont {H.~J.}\ \bibnamefont {Trodahl}}, \bibinfo {author}
  {\bibfnamefont {N.~M.}\ \bibnamefont {Strickland}}, \bibinfo {author}
  {\bibfnamefont {A.}~\bibnamefont {Koo}}, \bibinfo {author} {\bibfnamefont
  {V.}~\bibnamefont {Goian}}, \ and\ \bibinfo {author} {\bibfnamefont
  {S.}~\bibnamefont {Kamba}},\ }\href {\doibase
  http://dx.doi.org/10.1063/1.4807647} {\bibfield  {journal} {\bibinfo
  {journal} {J. Appl. Phys.}\ }\textbf {\bibinfo {volume} {113}},\ \bibinfo
  {eid} {203509} (\bibinfo {year} {2013})}\BibitemShut {NoStop}%
\bibitem [{\citenamefont {Warring}\ \emph {et~al.}(2014)\citenamefont
  {Warring}, \citenamefont {Ruck}, \citenamefont {McNulty}, \citenamefont
  {Anton}, \citenamefont {Granville}, \citenamefont {Koo}, \citenamefont
  {Cowie},\ and\ \citenamefont {Trodahl}}]{Warring_PRB_2014}%
  \BibitemOpen
  \bibfield  {author} {\bibinfo {author} {\bibfnamefont {H.}~\bibnamefont
  {Warring}}, \bibinfo {author} {\bibfnamefont {B.~J.}\ \bibnamefont {Ruck}},
  \bibinfo {author} {\bibfnamefont {J.~F.}\ \bibnamefont {McNulty}}, \bibinfo
  {author} {\bibfnamefont {E.-M.}\ \bibnamefont {Anton}}, \bibinfo {author}
  {\bibfnamefont {S.}~\bibnamefont {Granville}}, \bibinfo {author}
  {\bibfnamefont {A.}~\bibnamefont {Koo}}, \bibinfo {author} {\bibfnamefont
  {B.}~\bibnamefont {Cowie}}, \ and\ \bibinfo {author} {\bibfnamefont {H.~J.}\
  \bibnamefont {Trodahl}},\ }\href {\doibase 10.1103/PhysRevB.90.245206}
  {\bibfield  {journal} {\bibinfo  {journal} {Phys. Rev. B}\ }\textbf {\bibinfo
  {volume} {90}},\ \bibinfo {pages} {245206} (\bibinfo {year}
  {2014})}\BibitemShut {NoStop}%
\bibitem [{\citenamefont {Senapati}\ \emph {et~al.}(2011)\citenamefont
  {Senapati}, \citenamefont {Blamire},\ and\ \citenamefont
  {Barber}}]{Senapati_Nature_2011}%
  \BibitemOpen
  \bibfield  {author} {\bibinfo {author} {\bibfnamefont {K.}~\bibnamefont
  {Senapati}}, \bibinfo {author} {\bibfnamefont {M.~G.}\ \bibnamefont
  {Blamire}}, \ and\ \bibinfo {author} {\bibfnamefont {Z.~H.}\ \bibnamefont
  {Barber}},\ }\href
  {http://www.nature.com/nmat/journal/v10/n11/abs/nmat3116.html} {\bibfield
  {journal} {\bibinfo  {journal} {Nat. Mater.}\ }\textbf {\bibinfo {volume}
  {10}},\ \bibinfo {pages} {849} (\bibinfo {year} {2011})}\BibitemShut
  {NoStop}%
\bibitem [{\citenamefont {Massarotti}\ \emph {et~al.}(2015)\citenamefont
  {Massarotti}, \citenamefont {Pal}, \citenamefont {Rotoli}, \citenamefont
  {Longobardi}, \citenamefont {Blamire},\ and\ \citenamefont
  {Tafuri}}]{Massarotti_NC_2015}%
  \BibitemOpen
  \bibfield  {author} {\bibinfo {author} {\bibfnamefont {D.}~\bibnamefont
  {Massarotti}}, \bibinfo {author} {\bibfnamefont {A.}~\bibnamefont {Pal}},
  \bibinfo {author} {\bibfnamefont {G.}~\bibnamefont {Rotoli}}, \bibinfo
  {author} {\bibfnamefont {L.}~\bibnamefont {Longobardi}}, \bibinfo {author}
  {\bibfnamefont {M.}~\bibnamefont {Blamire}}, \ and\ \bibinfo {author}
  {\bibfnamefont {F.}~\bibnamefont {Tafuri}},\ }\href@noop {} {\bibfield
  {journal} {\bibinfo  {journal} {Nat. Commun.}\ }\textbf {\bibinfo {volume}
  {6}} (\bibinfo {year} {2015})}\BibitemShut {NoStop}%
\bibitem [{\citenamefont {Warring}\ \emph {et~al.}(2013)\citenamefont
  {Warring}, \citenamefont {Ruck}, \citenamefont {Trodahl},\ and\ \citenamefont
  {Natali}}]{Warring_APL_2013}%
  \BibitemOpen
  \bibfield  {author} {\bibinfo {author} {\bibfnamefont {H.}~\bibnamefont
  {Warring}}, \bibinfo {author} {\bibfnamefont {B.}~\bibnamefont {Ruck}},
  \bibinfo {author} {\bibfnamefont {H.}~\bibnamefont {Trodahl}}, \ and\
  \bibinfo {author} {\bibfnamefont {F.}~\bibnamefont {Natali}},\ }\href@noop {}
  {\bibfield  {journal} {\bibinfo  {journal} {Appl. Phys. Lett.}\ }\textbf
  {\bibinfo {volume} {102}},\ \bibinfo {pages} {132409} (\bibinfo {year}
  {2013})}\BibitemShut {NoStop}%
\bibitem [{\citenamefont {Muduli}\ \emph {et~al.}(2014)\citenamefont {Muduli},
  \citenamefont {Pal},\ and\ \citenamefont {Blamire}}]{Muduli_PRB_2014}%
  \BibitemOpen
  \bibfield  {author} {\bibinfo {author} {\bibfnamefont {P.~K.}\ \bibnamefont
  {Muduli}}, \bibinfo {author} {\bibfnamefont {A.}~\bibnamefont {Pal}}, \ and\
  \bibinfo {author} {\bibfnamefont {M.~G.}\ \bibnamefont {Blamire}},\ }\href
  {\doibase 10.1103/PhysRevB.89.094414} {\bibfield  {journal} {\bibinfo
  {journal} {Phys. Rev. B}\ }\textbf {\bibinfo {volume} {89}},\ \bibinfo
  {pages} {094414} (\bibinfo {year} {2014})}\BibitemShut {NoStop}%
\bibitem [{\citenamefont {McNulty}\ \emph {et~al.}(2015)\citenamefont
  {McNulty}, \citenamefont {Anton}, \citenamefont {Ruck}, \citenamefont
  {Natali}, \citenamefont {Warring}, \citenamefont {Wilhelm}, \citenamefont
  {Rogalev}, \citenamefont {Soares}, \citenamefont {Brookes},\ and\
  \citenamefont {Trodahl}}]{McNulty_PRB_2015}%
  \BibitemOpen
  \bibfield  {author} {\bibinfo {author} {\bibfnamefont {J.~F.}\ \bibnamefont
  {McNulty}}, \bibinfo {author} {\bibfnamefont {E.-M.}\ \bibnamefont {Anton}},
  \bibinfo {author} {\bibfnamefont {B.~J.}\ \bibnamefont {Ruck}}, \bibinfo
  {author} {\bibfnamefont {F.}~\bibnamefont {Natali}}, \bibinfo {author}
  {\bibfnamefont {H.}~\bibnamefont {Warring}}, \bibinfo {author} {\bibfnamefont
  {F.}~\bibnamefont {Wilhelm}}, \bibinfo {author} {\bibfnamefont
  {A.}~\bibnamefont {Rogalev}}, \bibinfo {author} {\bibfnamefont {M.~M.}\
  \bibnamefont {Soares}}, \bibinfo {author} {\bibfnamefont {N.~B.}\
  \bibnamefont {Brookes}}, \ and\ \bibinfo {author} {\bibfnamefont {H.~J.}\
  \bibnamefont {Trodahl}},\ }\href {\doibase 10.1103/PhysRevB.91.174426}
  {\bibfield  {journal} {\bibinfo  {journal} {Phys. Rev. B}\ }\textbf {\bibinfo
  {volume} {91}},\ \bibinfo {pages} {174426} (\bibinfo {year}
  {2015})}\BibitemShut {NoStop}%
\bibitem [{\citenamefont {Anton}\ \emph {et~al.}(2015)\citenamefont {Anton},
  \citenamefont {Granville}, \citenamefont {Engel}, \citenamefont {Chong},
  \citenamefont {Governale}, \citenamefont {Z{\"u}licke}, \citenamefont
  {Moghaddam}, \citenamefont {Trodahl}, \citenamefont {Natali}, \citenamefont
  {V{\'e}zian},\ and\ \citenamefont {Ruck}}]{Anton_Arxiv_2015}%
  \BibitemOpen
  \bibfield  {author} {\bibinfo {author} {\bibfnamefont {E.-M.}\ \bibnamefont
  {Anton}}, \bibinfo {author} {\bibfnamefont {S.}~\bibnamefont {Granville}},
  \bibinfo {author} {\bibfnamefont {A.}~\bibnamefont {Engel}}, \bibinfo
  {author} {\bibfnamefont {S.~V.}\ \bibnamefont {Chong}}, \bibinfo {author}
  {\bibfnamefont {M.}~\bibnamefont {Governale}}, \bibinfo {author}
  {\bibfnamefont {U.}~\bibnamefont {Z{\"u}licke}}, \bibinfo {author}
  {\bibfnamefont {A.~G.}\ \bibnamefont {Moghaddam}}, \bibinfo {author}
  {\bibfnamefont {H.~J.}\ \bibnamefont {Trodahl}}, \bibinfo {author}
  {\bibfnamefont {F.}~\bibnamefont {Natali}}, \bibinfo {author} {\bibfnamefont
  {S.}~\bibnamefont {V{\'e}zian}}, \ and\ \bibinfo {author} {\bibfnamefont
  {B.~J.}\ \bibnamefont {Ruck}},\ }\href@noop {} {\  (\bibinfo {year}
  {2015})},\ \Eprint {http://arxiv.org/abs/1505.03621} {arXiv:1505.03621
  [cond-mat]} \BibitemShut {NoStop}%
\bibitem [{\citenamefont {Anton}\ \emph {et~al.}(2013)\citenamefont {Anton},
  \citenamefont {Ruck}, \citenamefont {Meyer}, \citenamefont {Natali},
  \citenamefont {Warring}, \citenamefont {Wilhelm}, \citenamefont {Rogalev},
  \citenamefont {Antonov},\ and\ \citenamefont {Trodahl}}]{Anton_PRB_2013}%
  \BibitemOpen
  \bibfield  {author} {\bibinfo {author} {\bibfnamefont {E.-M.}\ \bibnamefont
  {Anton}}, \bibinfo {author} {\bibfnamefont {B.~J.}\ \bibnamefont {Ruck}},
  \bibinfo {author} {\bibfnamefont {C.}~\bibnamefont {Meyer}}, \bibinfo
  {author} {\bibfnamefont {F.}~\bibnamefont {Natali}}, \bibinfo {author}
  {\bibfnamefont {H.}~\bibnamefont {Warring}}, \bibinfo {author} {\bibfnamefont
  {F.}~\bibnamefont {Wilhelm}}, \bibinfo {author} {\bibfnamefont
  {A.}~\bibnamefont {Rogalev}}, \bibinfo {author} {\bibfnamefont {V.~N.}\
  \bibnamefont {Antonov}}, \ and\ \bibinfo {author} {\bibfnamefont {H.~J.}\
  \bibnamefont {Trodahl}},\ }\href {\doibase 10.1103/PhysRevB.87.134414}
  {\bibfield  {journal} {\bibinfo  {journal} {Phys. Rev. B}\ }\textbf {\bibinfo
  {volume} {87}},\ \bibinfo {pages} {134414} (\bibinfo {year}
  {2013})}\BibitemShut {NoStop}%
\bibitem [{\citenamefont {Ruck}\ \emph {et~al.}(2011)\citenamefont {Ruck},
  \citenamefont {Trodahl}, \citenamefont {Richter}, \citenamefont {Cezar},
  \citenamefont {Wilhelm}, \citenamefont {Rogalev}, \citenamefont {Antonov},
  \citenamefont {Le},\ and\ \citenamefont {Meyer}}]{Ruck_PRB_2011}%
  \BibitemOpen
  \bibfield  {author} {\bibinfo {author} {\bibfnamefont {B.~J.}\ \bibnamefont
  {Ruck}}, \bibinfo {author} {\bibfnamefont {H.~J.}\ \bibnamefont {Trodahl}},
  \bibinfo {author} {\bibfnamefont {J.~H.}\ \bibnamefont {Richter}}, \bibinfo
  {author} {\bibfnamefont {J.~C.}\ \bibnamefont {Cezar}}, \bibinfo {author}
  {\bibfnamefont {F.}~\bibnamefont {Wilhelm}}, \bibinfo {author} {\bibfnamefont
  {A.}~\bibnamefont {Rogalev}}, \bibinfo {author} {\bibfnamefont {V.~N.}\
  \bibnamefont {Antonov}}, \bibinfo {author} {\bibfnamefont {B.~D.}\
  \bibnamefont {Le}}, \ and\ \bibinfo {author} {\bibfnamefont {C.}~\bibnamefont
  {Meyer}},\ }\href {\doibase 10.1103/PhysRevB.83.174404} {\bibfield  {journal}
  {\bibinfo  {journal} {Phys. Rev. B}\ }\textbf {\bibinfo {volume} {83}},\
  \bibinfo {pages} {174404} (\bibinfo {year} {2011})}\BibitemShut {NoStop}%
\bibitem [{\citenamefont {De~Wijn}\ \emph {et~al.}(1976)\citenamefont
  {De~Wijn}, \citenamefont {Van~Diepen},\ and\ \citenamefont
  {Buschow}}]{DeWijn_PSS_1976}%
  \BibitemOpen
  \bibfield  {author} {\bibinfo {author} {\bibfnamefont {H.}~\bibnamefont
  {De~Wijn}}, \bibinfo {author} {\bibfnamefont {A.}~\bibnamefont {Van~Diepen}},
  \ and\ \bibinfo {author} {\bibfnamefont {K.}~\bibnamefont {Buschow}},\
  }\href@noop {} {\bibfield  {journal} {\bibinfo  {journal} {Phys. Status
  Solidi (B)}\ }\textbf {\bibinfo {volume} {76}},\ \bibinfo {pages} {11}
  (\bibinfo {year} {1976})}\BibitemShut {NoStop}%
\bibitem [{\citenamefont {Adachi}\ \emph {et~al.}(1997)\citenamefont {Adachi},
  \citenamefont {Ino},\ and\ \citenamefont {Miwa}}]{Adachi_PRB_1997}%
  \BibitemOpen
  \bibfield  {author} {\bibinfo {author} {\bibfnamefont {H.}~\bibnamefont
  {Adachi}}, \bibinfo {author} {\bibfnamefont {H.}~\bibnamefont {Ino}}, \ and\
  \bibinfo {author} {\bibfnamefont {H.}~\bibnamefont {Miwa}},\ }\href {\doibase
  10.1103/PhysRevB.56.349} {\bibfield  {journal} {\bibinfo  {journal} {Phys.
  Rev. B}\ }\textbf {\bibinfo {volume} {56}},\ \bibinfo {pages} {349} (\bibinfo
  {year} {1997})}\BibitemShut {NoStop}%
\bibitem [{\citenamefont {Mitra}\ and\ \citenamefont
  {Lambrecht}(2008)}]{Mitra_PRB_2008}%
  \BibitemOpen
  \bibfield  {author} {\bibinfo {author} {\bibfnamefont {C.}~\bibnamefont
  {Mitra}}\ and\ \bibinfo {author} {\bibfnamefont {W.~R.~L.}\ \bibnamefont
  {Lambrecht}},\ }\href {\doibase 10.1103/PhysRevB.78.134421} {\bibfield
  {journal} {\bibinfo  {journal} {Phys. Rev. B}\ }\textbf {\bibinfo {volume}
  {78}},\ \bibinfo {pages} {134421} (\bibinfo {year} {2008})}\BibitemShut
  {NoStop}%
\bibitem [{\citenamefont {Sharma}\ and\ \citenamefont
  {Nolting}(2010)}]{Sharma_PRB_2010}%
  \BibitemOpen
  \bibfield  {author} {\bibinfo {author} {\bibfnamefont {A.}~\bibnamefont
  {Sharma}}\ and\ \bibinfo {author} {\bibfnamefont {W.}~\bibnamefont
  {Nolting}},\ }\href {\doibase 10.1103/PhysRevB.81.125303} {\bibfield
  {journal} {\bibinfo  {journal} {Phys. Rev. B}\ }\textbf {\bibinfo {volume}
  {81}},\ \bibinfo {pages} {125303} (\bibinfo {year} {2010})}\BibitemShut
  {NoStop}%
\bibitem [{\citenamefont {Natali}\ \emph
  {et~al.}(2013{\natexlab{b}})\citenamefont {Natali}, \citenamefont {Ruck},
  \citenamefont {Trodahl}, \citenamefont {Binh}, \citenamefont {Vezian},
  \citenamefont {Damilano}, \citenamefont {Cordier}, \citenamefont {Semond},\
  and\ \citenamefont {Meyer}}]{Natali_PRB_2013}%
  \BibitemOpen
  \bibfield  {author} {\bibinfo {author} {\bibfnamefont {F.}~\bibnamefont
  {Natali}}, \bibinfo {author} {\bibfnamefont {B.~J.}\ \bibnamefont {Ruck}},
  \bibinfo {author} {\bibfnamefont {H.~J.}\ \bibnamefont {Trodahl}}, \bibinfo
  {author} {\bibfnamefont {D.~L.}\ \bibnamefont {Binh}}, \bibinfo {author}
  {\bibfnamefont {S.}~\bibnamefont {Vezian}}, \bibinfo {author} {\bibfnamefont
  {B.}~\bibnamefont {Damilano}}, \bibinfo {author} {\bibfnamefont
  {Y.}~\bibnamefont {Cordier}}, \bibinfo {author} {\bibfnamefont
  {F.}~\bibnamefont {Semond}}, \ and\ \bibinfo {author} {\bibfnamefont
  {C.}~\bibnamefont {Meyer}},\ }\href {\doibase 10.1103/PhysRevB.87.035202}
  {\bibfield  {journal} {\bibinfo  {journal} {Phys. Rev. B}\ }\textbf {\bibinfo
  {volume} {87}},\ \bibinfo {pages} {035202} (\bibinfo {year}
  {2013}{\natexlab{b}})}\BibitemShut {NoStop}%
\bibitem [{\citenamefont {Lee}\ \emph {et~al.}(2015)\citenamefont {Lee},
  \citenamefont {Warring}, \citenamefont {V{\'e}zian}, \citenamefont
  {Damilano}, \citenamefont {Granville}, \citenamefont {Al~Khalfioui},
  \citenamefont {Cordier}, \citenamefont {Trodahl}, \citenamefont {Ruck},\ and\
  \citenamefont {Natali}}]{Lee_APL_2015}%
  \BibitemOpen
  \bibfield  {author} {\bibinfo {author} {\bibfnamefont {C.-M.}\ \bibnamefont
  {Lee}}, \bibinfo {author} {\bibfnamefont {H.}~\bibnamefont {Warring}},
  \bibinfo {author} {\bibfnamefont {S.}~\bibnamefont {V{\'e}zian}}, \bibinfo
  {author} {\bibfnamefont {B.}~\bibnamefont {Damilano}}, \bibinfo {author}
  {\bibfnamefont {S.}~\bibnamefont {Granville}}, \bibinfo {author}
  {\bibfnamefont {M.}~\bibnamefont {Al~Khalfioui}}, \bibinfo {author}
  {\bibfnamefont {Y.}~\bibnamefont {Cordier}}, \bibinfo {author} {\bibfnamefont
  {H.~J.}\ \bibnamefont {Trodahl}}, \bibinfo {author} {\bibfnamefont {B.~J.}\
  \bibnamefont {Ruck}}, \ and\ \bibinfo {author} {\bibfnamefont
  {F.}~\bibnamefont {Natali}},\ }\href@noop {} {\bibfield  {journal} {\bibinfo
  {journal} {Appl. Phys. Lett.}\ }\textbf {\bibinfo {volume} {106}},\ \bibinfo
  {pages} {022401} (\bibinfo {year} {2015})}\BibitemShut {NoStop}%
\bibitem [{\citenamefont {Larson}\ \emph {et~al.}(2007)\citenamefont {Larson},
  \citenamefont {Lambrecht}, \citenamefont {Chantis},\ and\ \citenamefont {van
  Schilfgaarde}}]{Larson_PRB_2007}%
  \BibitemOpen
  \bibfield  {author} {\bibinfo {author} {\bibfnamefont {P.}~\bibnamefont
  {Larson}}, \bibinfo {author} {\bibfnamefont {W.~R.~L.}\ \bibnamefont
  {Lambrecht}}, \bibinfo {author} {\bibfnamefont {A.}~\bibnamefont {Chantis}},
  \ and\ \bibinfo {author} {\bibfnamefont {M.}~\bibnamefont {van
  Schilfgaarde}},\ }\href {\doibase 10.1103/PhysRevB.75.045114} {\bibfield
  {journal} {\bibinfo  {journal} {Phys. Rev. B}\ }\textbf {\bibinfo {volume}
  {75}},\ \bibinfo {pages} {045114} (\bibinfo {year} {2007})}\BibitemShut
  {NoStop}%
\bibitem [{\citenamefont {Morari}\ \emph {et~al.}(2015)\citenamefont {Morari},
  \citenamefont {Beiu{\c{s}}eanu}, \citenamefont {Di~Marco}, \citenamefont
  {Peters}, \citenamefont {Burzo}, \citenamefont {Mican},\ and\ \citenamefont
  {Chioncel}}]{Morari_JPCM_2014}%
  \BibitemOpen
  \bibfield  {author} {\bibinfo {author} {\bibfnamefont {C.}~\bibnamefont
  {Morari}}, \bibinfo {author} {\bibfnamefont {F.}~\bibnamefont
  {Beiu{\c{s}}eanu}}, \bibinfo {author} {\bibfnamefont {I.}~\bibnamefont
  {Di~Marco}}, \bibinfo {author} {\bibfnamefont {L.}~\bibnamefont {Peters}},
  \bibinfo {author} {\bibfnamefont {E.}~\bibnamefont {Burzo}}, \bibinfo
  {author} {\bibfnamefont {S.}~\bibnamefont {Mican}}, \ and\ \bibinfo {author}
  {\bibfnamefont {L.}~\bibnamefont {Chioncel}},\ }\href
  {http://iopscience.iop.org/0953-8984/27/11/115503} {\bibfield  {journal}
  {\bibinfo  {journal} {J. Phys.: Condens. Matter}\ }\textbf {\bibinfo {volume}
  {27}},\ \bibinfo {pages} {115503} (\bibinfo {year} {2015})}\BibitemShut
  {NoStop}%
\bibitem [{\citenamefont {Peters}\ \emph {et~al.}(2014)\citenamefont {Peters},
  \citenamefont {Di~Marco}, \citenamefont {Thunstr\"om}, \citenamefont
  {Katsnelson}, \citenamefont {Kirilyuk},\ and\ \citenamefont
  {Eriksson}}]{Peters_PRB_2014}%
  \BibitemOpen
  \bibfield  {author} {\bibinfo {author} {\bibfnamefont {L.}~\bibnamefont
  {Peters}}, \bibinfo {author} {\bibfnamefont {I.}~\bibnamefont {Di~Marco}},
  \bibinfo {author} {\bibfnamefont {P.}~\bibnamefont {Thunstr\"om}}, \bibinfo
  {author} {\bibfnamefont {M.~I.}\ \bibnamefont {Katsnelson}}, \bibinfo
  {author} {\bibfnamefont {A.}~\bibnamefont {Kirilyuk}}, \ and\ \bibinfo
  {author} {\bibfnamefont {O.}~\bibnamefont {Eriksson}},\ }\href {\doibase
  10.1103/PhysRevB.89.205109} {\bibfield  {journal} {\bibinfo  {journal} {Phys.
  Rev. B}\ }\textbf {\bibinfo {volume} {89}},\ \bibinfo {pages} {205109}
  (\bibinfo {year} {2014})}\BibitemShut {NoStop}%
\bibitem [{\citenamefont {van Diepen}\ \emph {et~al.}(1973)\citenamefont {van
  Diepen}, \citenamefont {de~Wijn},\ and\ \citenamefont
  {Buschow}}]{Diepen_PRB_1973}%
  \BibitemOpen
  \bibfield  {author} {\bibinfo {author} {\bibfnamefont {A.~M.}\ \bibnamefont
  {van Diepen}}, \bibinfo {author} {\bibfnamefont {H.~W.}\ \bibnamefont
  {de~Wijn}}, \ and\ \bibinfo {author} {\bibfnamefont {K.~H.~J.}\ \bibnamefont
  {Buschow}},\ }\href {\doibase 10.1103/PhysRevB.8.1125} {\bibfield  {journal}
  {\bibinfo  {journal} {Phys. Rev. B}\ }\textbf {\bibinfo {volume} {8}},\
  \bibinfo {pages} {1125} (\bibinfo {year} {1973})}\BibitemShut {NoStop}%
\bibitem [{\citenamefont {Buschow}\ \emph {et~al.}(1973)\citenamefont
  {Buschow}, \citenamefont {van Diepen},\ and\ \citenamefont
  {de~Wijn}}]{Buschow_PRB_1973}%
  \BibitemOpen
  \bibfield  {author} {\bibinfo {author} {\bibfnamefont {K.~H.~J.}\
  \bibnamefont {Buschow}}, \bibinfo {author} {\bibfnamefont {A.~M.}\
  \bibnamefont {van Diepen}}, \ and\ \bibinfo {author} {\bibfnamefont {H.~W.}\
  \bibnamefont {de~Wijn}},\ }\href {\doibase 10.1103/PhysRevB.8.5134}
  {\bibfield  {journal} {\bibinfo  {journal} {Phys. Rev. B}\ }\textbf {\bibinfo
  {volume} {8}},\ \bibinfo {pages} {5134} (\bibinfo {year} {1973})}\BibitemShut
  {NoStop}%
\bibitem [{\citenamefont {Adachi}\ and\ \citenamefont
  {Ino}(1999)}]{Adachi_Nature_1999}%
  \BibitemOpen
  \bibfield  {author} {\bibinfo {author} {\bibfnamefont {H.}~\bibnamefont
  {Adachi}}\ and\ \bibinfo {author} {\bibfnamefont {H.}~\bibnamefont {Ino}},\
  }\href {http://www.nature.com/nature/journal/v401/n6749/abs/401148a0.html}
  {\bibfield  {journal} {\bibinfo  {journal} {Nature}\ }\textbf {\bibinfo
  {volume} {401}},\ \bibinfo {pages} {148} (\bibinfo {year}
  {1999})}\BibitemShut {NoStop}%
\bibitem [{\citenamefont {Adachi}\ \emph {et~al.}(2001)\citenamefont {Adachi},
  \citenamefont {Kawata}, \citenamefont {Hashimoto}, \citenamefont {Sato},
  \citenamefont {Matsumoto},\ and\ \citenamefont {Tanaka}}]{Adachi_PRL_2001}%
  \BibitemOpen
  \bibfield  {author} {\bibinfo {author} {\bibfnamefont {H.}~\bibnamefont
  {Adachi}}, \bibinfo {author} {\bibfnamefont {H.}~\bibnamefont {Kawata}},
  \bibinfo {author} {\bibfnamefont {H.}~\bibnamefont {Hashimoto}}, \bibinfo
  {author} {\bibfnamefont {Y.}~\bibnamefont {Sato}}, \bibinfo {author}
  {\bibfnamefont {I.}~\bibnamefont {Matsumoto}}, \ and\ \bibinfo {author}
  {\bibfnamefont {Y.}~\bibnamefont {Tanaka}},\ }\href {\doibase
  10.1103/PhysRevLett.87.127202} {\bibfield  {journal} {\bibinfo  {journal}
  {Phys. Rev. Lett.}\ }\textbf {\bibinfo {volume} {87}},\ \bibinfo {pages}
  {127202} (\bibinfo {year} {2001})}\BibitemShut {NoStop}%
\bibitem [{\citenamefont {Qiao}\ \emph {et~al.}(2004)\citenamefont {Qiao},
  \citenamefont {Kimura}, \citenamefont {Adachi}, \citenamefont {Iori},
  \citenamefont {Miyamoto}, \citenamefont {Xie}, \citenamefont {Namatame},
  \citenamefont {Taniguchi}, \citenamefont {Tanaka}, \citenamefont {Muro},
  \citenamefont {Imada},\ and\ \citenamefont {Suga}}]{Qiao_PRB_2004}%
  \BibitemOpen
  \bibfield  {author} {\bibinfo {author} {\bibfnamefont {S.}~\bibnamefont
  {Qiao}}, \bibinfo {author} {\bibfnamefont {A.}~\bibnamefont {Kimura}},
  \bibinfo {author} {\bibfnamefont {H.}~\bibnamefont {Adachi}}, \bibinfo
  {author} {\bibfnamefont {K.}~\bibnamefont {Iori}}, \bibinfo {author}
  {\bibfnamefont {K.}~\bibnamefont {Miyamoto}}, \bibinfo {author}
  {\bibfnamefont {T.}~\bibnamefont {Xie}}, \bibinfo {author} {\bibfnamefont
  {H.}~\bibnamefont {Namatame}}, \bibinfo {author} {\bibfnamefont
  {M.}~\bibnamefont {Taniguchi}}, \bibinfo {author} {\bibfnamefont
  {A.}~\bibnamefont {Tanaka}}, \bibinfo {author} {\bibfnamefont
  {T.}~\bibnamefont {Muro}}, \bibinfo {author} {\bibfnamefont {S.}~\bibnamefont
  {Imada}}, \ and\ \bibinfo {author} {\bibfnamefont {S.}~\bibnamefont {Suga}},\
  }\href {\doibase 10.1103/PhysRevB.70.134418} {\bibfield  {journal} {\bibinfo
  {journal} {Phys. Rev. B}\ }\textbf {\bibinfo {volume} {70}},\ \bibinfo
  {pages} {134418} (\bibinfo {year} {2004})}\BibitemShut {NoStop}%
\bibitem [{\citenamefont {Dhesi}\ \emph {et~al.}(2010)\citenamefont {Dhesi},
  \citenamefont {van~der Laan}, \citenamefont {Bencok}, \citenamefont
  {Brookes}, \citenamefont {Gal\'era},\ and\ \citenamefont
  {Ohresser}}]{Dhesi_PRB_2010}%
  \BibitemOpen
  \bibfield  {author} {\bibinfo {author} {\bibfnamefont {S.~S.}\ \bibnamefont
  {Dhesi}}, \bibinfo {author} {\bibfnamefont {G.}~\bibnamefont {van~der Laan}},
  \bibinfo {author} {\bibfnamefont {P.}~\bibnamefont {Bencok}}, \bibinfo
  {author} {\bibfnamefont {N.~B.}\ \bibnamefont {Brookes}}, \bibinfo {author}
  {\bibfnamefont {R.~M.}\ \bibnamefont {Gal\'era}}, \ and\ \bibinfo {author}
  {\bibfnamefont {P.}~\bibnamefont {Ohresser}},\ }\href {\doibase
  10.1103/PhysRevB.82.180402} {\bibfield  {journal} {\bibinfo  {journal} {Phys.
  Rev. B}\ }\textbf {\bibinfo {volume} {82}},\ \bibinfo {pages} {180402}
  (\bibinfo {year} {2010})}\BibitemShut {NoStop}%
\bibitem [{\citenamefont {Adachi}\ \emph {et~al.}(1999)\citenamefont {Adachi},
  \citenamefont {Ino},\ and\ \citenamefont {Miwa}}]{Adachi_PRB_1999}%
  \BibitemOpen
  \bibfield  {author} {\bibinfo {author} {\bibfnamefont {H.}~\bibnamefont
  {Adachi}}, \bibinfo {author} {\bibfnamefont {H.}~\bibnamefont {Ino}}, \ and\
  \bibinfo {author} {\bibfnamefont {H.}~\bibnamefont {Miwa}},\ }\href {\doibase
  10.1103/PhysRevB.59.11445} {\bibfield  {journal} {\bibinfo  {journal} {Phys.
  Rev. B}\ }\textbf {\bibinfo {volume} {59}},\ \bibinfo {pages} {11445}
  (\bibinfo {year} {1999})}\BibitemShut {NoStop}%
\bibitem [{\citenamefont {Stevens}(1952)}]{Stevens_PPSA_1952}%
  \BibitemOpen
  \bibfield  {author} {\bibinfo {author} {\bibfnamefont {K.}~\bibnamefont
  {Stevens}},\ }\href@noop {} {\bibfield  {journal} {\bibinfo  {journal} {Proc.
  Phys. Soc. London, Sect. A}\ }\textbf {\bibinfo {volume} {65}},\ \bibinfo
  {pages} {209} (\bibinfo {year} {1952})}\BibitemShut {NoStop}%
\bibitem [{\citenamefont {Hutchings}(1964)}]{Hutchings_SSP_1964}%
  \BibitemOpen
  \bibfield  {author} {\bibinfo {author} {\bibfnamefont {M.~T.}\ \bibnamefont
  {Hutchings}},\ }\href@noop {} {\bibfield  {journal} {\bibinfo  {journal}
  {Solid State Physics}\ }\textbf {\bibinfo {volume} {16}},\ \bibinfo {pages}
  {227} (\bibinfo {year} {1964})}\BibitemShut {NoStop}%
\bibitem [{Note1()}]{Note1}%
  \BibitemOpen
  \bibinfo {note} {We have found $z$ parallel to $[111]$ to be the lowest
  energy orientation compared to $[001]$ and $[110]$ directions. This was also
  reported in \cite {Buschow_PRB_1973} and \cite {Adachi_PRB_1999} for
  Sm$^{3+}$}\BibitemShut {NoStop}%
\bibitem [{\citenamefont {Birgeneau}\ \emph {et~al.}(1973)\citenamefont
  {Birgeneau}, \citenamefont {Bucher}, \citenamefont {Maita}, \citenamefont
  {Passell},\ and\ \citenamefont {Turberfield}}]{Birgeneau_PRB_1973}%
  \BibitemOpen
  \bibfield  {author} {\bibinfo {author} {\bibfnamefont {R.~J.}\ \bibnamefont
  {Birgeneau}}, \bibinfo {author} {\bibfnamefont {E.}~\bibnamefont {Bucher}},
  \bibinfo {author} {\bibfnamefont {J.~P.}\ \bibnamefont {Maita}}, \bibinfo
  {author} {\bibfnamefont {L.}~\bibnamefont {Passell}}, \ and\ \bibinfo
  {author} {\bibfnamefont {K.~C.}\ \bibnamefont {Turberfield}},\ }\href
  {\doibase 10.1103/PhysRevB.8.5345} {\bibfield  {journal} {\bibinfo  {journal}
  {Phys. Rev. B}\ }\textbf {\bibinfo {volume} {8}},\ \bibinfo {pages} {5345}
  (\bibinfo {year} {1973})}\BibitemShut {NoStop}%
\bibitem [{\citenamefont {Hulliger}(1979)}]{Hulliger_HPCRE_1979}%
  \BibitemOpen
  \bibfield  {author} {\bibinfo {author} {\bibfnamefont {F.}~\bibnamefont
  {Hulliger}},\ }\href@noop {} {\bibfield  {journal} {\bibinfo  {journal}
  {Handbook on the Physics and Chemistry of Rare Earths}\ }\textbf {\bibinfo
  {volume} {4}},\ \bibinfo {pages} {153} (\bibinfo {year} {1979})}\BibitemShut
  {NoStop}%
\bibitem [{\citenamefont {Lea}\ \emph {et~al.}(1962)\citenamefont {Lea},
  \citenamefont {Leask},\ and\ \citenamefont {Wolf}}]{Lea_JPCS_1962}%
  \BibitemOpen
  \bibfield  {author} {\bibinfo {author} {\bibfnamefont {K.}~\bibnamefont
  {Lea}}, \bibinfo {author} {\bibfnamefont {M.}~\bibnamefont {Leask}}, \ and\
  \bibinfo {author} {\bibfnamefont {W.}~\bibnamefont {Wolf}},\ }\href@noop {}
  {\bibfield  {journal} {\bibinfo  {journal} {J. Phys. Chem. Solids}\ }\textbf
  {\bibinfo {volume} {23}},\ \bibinfo {pages} {1381} (\bibinfo {year}
  {1962})}\BibitemShut {NoStop}%
\bibitem [{\citenamefont {Freeman}\ and\ \citenamefont
  {Watson}(1962)}]{Freeman_PR_1962}%
  \BibitemOpen
  \bibfield  {author} {\bibinfo {author} {\bibfnamefont {A.~J.}\ \bibnamefont
  {Freeman}}\ and\ \bibinfo {author} {\bibfnamefont {R.~E.}\ \bibnamefont
  {Watson}},\ }\href {\doibase 10.1103/PhysRev.127.2058} {\bibfield  {journal}
  {\bibinfo  {journal} {Phys. Rev.}\ }\textbf {\bibinfo {volume} {127}},\
  \bibinfo {pages} {2058} (\bibinfo {year} {1962})}\BibitemShut {NoStop}%
\bibitem [{\citenamefont {Freeman}\ and\ \citenamefont
  {Desclaux}(1979)}]{Freeman_JMMM_1979}%
  \BibitemOpen
  \bibfield  {author} {\bibinfo {author} {\bibfnamefont {A.}~\bibnamefont
  {Freeman}}\ and\ \bibinfo {author} {\bibfnamefont {J.}~\bibnamefont
  {Desclaux}},\ }\href@noop {} {\bibfield  {journal} {\bibinfo  {journal} {J.
  Magn. Magn. Mater.}\ }\textbf {\bibinfo {volume} {12}},\ \bibinfo {pages}
  {11} (\bibinfo {year} {1979})}\BibitemShut {NoStop}%
\bibitem [{Note2()}]{Note2}%
  \BibitemOpen
  \bibinfo {note} {To less than 1\% error one may scale both $\protect
  \ensuremath {\left \delimiter "426830A r^4\right \delimiter "526930B }$ and
  $\protect \ensuremath {\left \delimiter "426830A r^6\right \delimiter
  "526930B }$ by 1.2 to go from Hartree-Fock to Dirac-Fock values, which is
  equivalent to rescaling $Z$. One must therefore scale $Z$ accordingly when
  comparing to values from the literature.}\BibitemShut {Stop}%
\bibitem [{\citenamefont {Thole}\ \emph {et~al.}(1992)\citenamefont {Thole},
  \citenamefont {Carra}, \citenamefont {Sette},\ and\ \citenamefont {van~der
  Laan}}]{Thole_PRL_1992}%
  \BibitemOpen
  \bibfield  {author} {\bibinfo {author} {\bibfnamefont {B.~T.}\ \bibnamefont
  {Thole}}, \bibinfo {author} {\bibfnamefont {P.}~\bibnamefont {Carra}},
  \bibinfo {author} {\bibfnamefont {F.}~\bibnamefont {Sette}}, \ and\ \bibinfo
  {author} {\bibfnamefont {G.}~\bibnamefont {van~der Laan}},\ }\href {\doibase
  10.1103/PhysRevLett.68.1943} {\bibfield  {journal} {\bibinfo  {journal}
  {Phys. Rev. Lett.}\ }\textbf {\bibinfo {volume} {68}},\ \bibinfo {pages}
  {1943} (\bibinfo {year} {1992})}\BibitemShut {NoStop}%
\bibitem [{\citenamefont {Carra}\ \emph
  {et~al.}(1993{\natexlab{a}})\citenamefont {Carra}, \citenamefont {Thole},
  \citenamefont {Altarelli},\ and\ \citenamefont {Wang}}]{Carra_PRL_1993}%
  \BibitemOpen
  \bibfield  {author} {\bibinfo {author} {\bibfnamefont {P.}~\bibnamefont
  {Carra}}, \bibinfo {author} {\bibfnamefont {B.~T.}\ \bibnamefont {Thole}},
  \bibinfo {author} {\bibfnamefont {M.}~\bibnamefont {Altarelli}}, \ and\
  \bibinfo {author} {\bibfnamefont {X.}~\bibnamefont {Wang}},\ }\href {\doibase
  10.1103/PhysRevLett.70.694} {\bibfield  {journal} {\bibinfo  {journal} {Phys.
  Rev. Lett.}\ }\textbf {\bibinfo {volume} {70}},\ \bibinfo {pages} {694}
  (\bibinfo {year} {1993}{\natexlab{a}})}\BibitemShut {NoStop}%
\bibitem [{\citenamefont {Carra}\ \emph
  {et~al.}(1993{\natexlab{b}})\citenamefont {Carra}, \citenamefont {K{\"o}nig},
  \citenamefont {Thole},\ and\ \citenamefont {Altarelli}}]{Carra_PB_1993}%
  \BibitemOpen
  \bibfield  {author} {\bibinfo {author} {\bibfnamefont {P.}~\bibnamefont
  {Carra}}, \bibinfo {author} {\bibfnamefont {H.}~\bibnamefont {K{\"o}nig}},
  \bibinfo {author} {\bibfnamefont {B.}~\bibnamefont {Thole}}, \ and\ \bibinfo
  {author} {\bibfnamefont {M.}~\bibnamefont {Altarelli}},\ }\href@noop {}
  {\bibfield  {journal} {\bibinfo  {journal} {Physica B: Condensed Matter}\
  }\textbf {\bibinfo {volume} {192}},\ \bibinfo {pages} {182} (\bibinfo {year}
  {1993}{\natexlab{b}})}\BibitemShut {NoStop}%
\bibitem [{\citenamefont {Cortie}\ \emph {et~al.}(2014)\citenamefont {Cortie},
  \citenamefont {Brown}, \citenamefont {Br\"uck}, \citenamefont {Saerbeck},
  \citenamefont {Evans}, \citenamefont {Fritzsche}, \citenamefont {Wang},
  \citenamefont {Downes},\ and\ \citenamefont {Klose}}]{Cortie_PRB_2014}%
  \BibitemOpen
  \bibfield  {author} {\bibinfo {author} {\bibfnamefont {D.~L.}\ \bibnamefont
  {Cortie}}, \bibinfo {author} {\bibfnamefont {J.~D.}\ \bibnamefont {Brown}},
  \bibinfo {author} {\bibfnamefont {S.}~\bibnamefont {Br\"uck}}, \bibinfo
  {author} {\bibfnamefont {T.}~\bibnamefont {Saerbeck}}, \bibinfo {author}
  {\bibfnamefont {J.~P.}\ \bibnamefont {Evans}}, \bibinfo {author}
  {\bibfnamefont {H.}~\bibnamefont {Fritzsche}}, \bibinfo {author}
  {\bibfnamefont {X.~L.}\ \bibnamefont {Wang}}, \bibinfo {author}
  {\bibfnamefont {J.~E.}\ \bibnamefont {Downes}}, \ and\ \bibinfo {author}
  {\bibfnamefont {F.}~\bibnamefont {Klose}},\ }\href {\doibase
  10.1103/PhysRevB.89.064424} {\bibfield  {journal} {\bibinfo  {journal} {Phys.
  Rev. B}\ }\textbf {\bibinfo {volume} {89}},\ \bibinfo {pages} {064424}
  (\bibinfo {year} {2014})}\BibitemShut {NoStop}%
\bibitem [{\citenamefont {Trammell}(1963)}]{Trammell_PR_1963}%
  \BibitemOpen
  \bibfield  {author} {\bibinfo {author} {\bibfnamefont {G.~T.}\ \bibnamefont
  {Trammell}},\ }\href {\doibase 10.1103/PhysRev.131.932} {\bibfield  {journal}
  {\bibinfo  {journal} {Phys. Rev.}\ }\textbf {\bibinfo {volume} {131}},\
  \bibinfo {pages} {932} (\bibinfo {year} {1963})}\BibitemShut {NoStop}%
\end{thebibliography}%

\end{document}